\documentclass{article}

\usepackage{arxiv}

\usepackage[utf8]{inputenc} 
\usepackage[T1]{fontenc}    
\usepackage{hyperref}       
\usepackage{url}            
\usepackage{booktabs}       
\usepackage{amsfonts}       
\usepackage{nicefrac}       
\usepackage{microtype}      
\usepackage{lipsum}
\usepackage{graphicx}
\usepackage[normalem]{ulem}
\usepackage{graphicx}%
\usepackage{multirow}%
\usepackage{amsmath,amssymb,amsfonts}%
\usepackage{amsthm}%
\usepackage{mathrsfs}%
\usepackage[title]{appendix}%
\usepackage{xcolor}%
\usepackage{epstopdf}
\usepackage{textcomp}%
\usepackage{manyfoot}%
\usepackage{booktabs}%
\usepackage{algorithm}%
\usepackage{algorithmicx}%
\usepackage{algpseudocode}%
\usepackage{listings}%
\usepackage{caption}
\usepackage{subcaption}

\title{Revisiting Conservativeness in Fluid Dynamics: Failure of Non-Conservative PINNs and a Path-Integral Remedy}

\author{
 Arun Govind Neelan \\
  SimuNetics\\
  Kanniyakumari \\
  Tamil Nadu\\
  India-629173\\
  \texttt{arunneelaniist@gmail.com} \\
   \And
 Ferdin Sagai Don Bosco \\
  BosonQ Psi (BQP)\\
  University of Pittsburgh\\
  New York \\
  USA-13202 \\
  \texttt{ferdinsagai7@gmail.com} \\
  \And
 Naveen Sagar Jarugumalli  \\
  Airbus India\\
  Bengaluru \\
  Karnataka\\
  India-560090
   \\
  \texttt{sagar.alex7@outlook.com} \\
   \AND
    Suresh Balaji Vedarethinam \\
   Indian Institute of Technology Kanpur \\
   Uttar Pradesh \\
   India-208016\\
   \texttt{vsbalaji15@gmail.com} \\
}

\begin{document}
\maketitle
\begin{abstract}
	The choice between conservative and non-conservative formulations is a fundamental dilemma in CFD. While non-conservative forms offer intuitive modeling in primitive variables, they typically produce erroneous shock speeds. This paper critically analyzes these formulations, contrasting classical failures against the capabilities of Physics-Informed Neural Networks (PINNs). 
	Using the \textbf{Adaptive Weight and Viscosity (PINNs-AWV)} architecture, we evaluate cases ranging from shallow water equations to unsteady 1D and 2D Euler equations. Results reveal a significant dichotomy: while PINNs-AWV restores physical fidelity in scalar and steady systems, standard non-conservative PINNs fail in unsteady systems like the Sod shock tube. We demonstrate this failure stems from non-vanishing source terms introduced by viscous regularization, which violate the Rankine--Hugoniot jump conditions. 		
	To resolve this, we implement a \textbf{path-integral framework} based on Dal Maso--LeFloch--Murat (DLM) theory. By incorporating path-consistent losses in PINNs (\textbf{PI-PINN}) and using path-conservative numerical schemes, we successfully recover correct shock speeds within non-conservative frameworks. Our results prove the path-integral approach provides a rigorous mathematical bridge for physical accuracy in both classical and machine learning solvers, enabling primitive-variable formulations in transient, high-speed simulations.
\end{abstract}


\section{Introduction}
The conservative and non-conservative forms are two ways of writing the mathematical model of a system as a partial differential equation (PDE). The conservative form of a governing equation expresses the conservation law in divergence (flux) form. Mathematically, it is written as:
\begin{equation}\label{eq:con}
	\frac{\partial \mathbf{U}}{\partial t} + \nabla \cdot \mathbf{F}(\mathbf{U}) = 0
\end{equation}
where, $\mathbf{U}$ vector of conserved variables, $\mathbf{F}$ is flux vector (which may depends on $\mathbf{U}$ for non-linear systems). Integrating \eqref{eq:con} over the cell of x = [i-0.5, i+0.5], we get conservative discretization 
\begin{equation}\label{eq:con1}
	\bar{\mathbf{U}}_i^{n+1} = \bar{\mathbf{U}}_i^n - \frac{\Delta t}{\Delta x_i} (\mathbf{F}_{i+1/2}^n - \mathbf{F}_{i-1/2}^n),
\end{equation}

where, $i$ is the cell center, and ${i+1/2}$ and ${i-1/2}$ are the cell faces.
Equation~\ref{eq:con1} is called conservative discretization. The numerical schemes based on the conservative formulations include the finite volume method~\cite{neelan2021three} and the conservative finite difference method~\cite{neelan2025higher}. In this work, reference to "conservative equations" will mean equations that can be written in the form of \eqref{eq:con}. 
Non-conservative form can be formulated by applying the chain rule to the conservative equation expressed in \eqref{eq:con}. For illustration, consider the law of mass conservation that is expressed in its conservative form as shown in eq.~\ref{eq.ContinuityEq_C_form}. Here, the flux of interest is $\rho \mathbf{u}$.

\begin{equation}\label{eq.ContinuityEq_C_form}
	\frac{\partial \rho}{\partial t} + \nabla \cdot (\rho \mathbf{u}) = 0
\end{equation}

On applying the product rule to the flux term, the resulting PDE can be expressed as, shown in eq.~\ref{eq.ContinuityEq_NC_form1} The non-conservative form of the continuity equation is

\begin{equation}\label{eq.ContinuityEq_NC_form1}
	\frac{D\rho}{Dt} + \rho (\nabla \cdot \mathbf{u}) = 0,
\end{equation}

where, $\frac{D\rho}{Dt}$ is the total/substantial derivative expressed as, $\frac{D\rho}{Dt} = \frac{\partial \rho}{\partial t} + \mathbf{u} . \nabla$, with $\mathbf{u} . \nabla \rho$ representing the convective derivative.
Mathematically, the conservative and non-conservative form are equivalent, with each being derivable from the other without any additional "mathematical" assumptions. 


These restrictions on the non-conservative formulation make it necessary to account for the physics of the system while selecting the form of the governing PDE. For instance, compressible flows with shocks is capable of accurately resolving the fluxes across the shock if the conservative form is utilized~\cite{toro2009riemann}. On the other hand, incompressible or smooth-flow regimes, where conservation across discontinuities is less critical, are more amenable to the non-conservation form of the equation. Additionally, this form operates directly on the primitive variables—velocity, pressure, and density—making it more intuitive than solutions from the conservative form, wherein the primitive variables need to be calculated from the fluxes. 

Application of the non-conservative form to compressible flows can be erroneous because this form lacks the ability to rigorously preserve conserved quantities across control volumes, leading to inaccurate solutions near shocks or discontinuities. Alternatively, conservative schemes tend to exhibit slower convergence for low Mach number flows. This is primarily because one of the eigenvalues of the flux Jacobian matrix—associated with the acoustic waves—approaches zero as the Mach number decreases~\cite{shapiro2006non}. Consequently, the system becomes stiff, leading to ill-conditioning in the numerical flux evaluations and degrading convergence performance. This can be overcome using pre-conditioners~\cite{guillard1999behavior},~\cite{turkel1999preconditioning}. 

{Rankine-Hugoniot relations ensure correct transport of mass, momentum, and energy across the discontinuity. The non-conservative form does not inherently satisfy these conditions.}
However, this comes at the cost of violating conservation laws, especially when strong gradients or shocks develop. 
It's important to reiterate that while non-conservative forms can converge faster for smooth, shock-free flows, their fundamental inability to correctly capture discontinuities without special treatment makes the conservative form the default choice for general-purpose Computational Fluid Dynamics (CFD) packages.
Though conservative schemes are widely accepted in CFD simulation packages, their implementation can sometimes reduce accuracy or efficiency on low-speed flows and the Schrodinger equation~\cite{verwer1986conservative}. 
The ``faster convergence" of the non-conservative form is a niche advantage, primarily relevant in specific applications where solution smoothness is guaranteed or shock-fitting techniques are employed.
The choice between conservative and non-conservative schemes should be based on the problem's nature, the importance of conservation, and the desired balance between accuracy, stability, and computational cost.
The summary of the advantages and disadvantages is tabulated in Table~\ref {tab:consv_nonconsv}.    

\begin{table}[h]
	\centering
	\caption{Comparison of conservative and non-conservative forms of governing equations}
	\label{tab:consv_nonconsv}
	\begin{tabular}{|l|c|c|}
		\hline
		\textbf{Aspect} & \textbf{Conservative Form} & \textbf{Non-Conservative Form} \\
		\hline
		High speed flows & Accurate shock prediction& may give wrong shocks  \\
		\hline
		Low speed flows & Slow convergence & Ideal \\
		\hline
		Discretization & FVM/FDM &  FDM \\
		\hline
		Computational cost & Slightly higher & Lower  \\
		\hline
		Conservation Property & Built-in & Mesh-dependent \\
		\hline
		
		Suitability & Compressible/Discontinuous flows & Smooth/Incompressible flows \\
		\hline
	\end{tabular}
\end{table}
While conservative formulations are ideal for problems involving shocks due to their ability to preserve physical quantities across discontinuities, there are many important cases where it is not possible to express the governing equations in a purely conservative form. A classic example is the shallow water equations (SWE): while they admit a conservative form over a flat bottom, introducing variable bottom topography breaks the conservative structure~\cite{li2024path},~\cite{LeVeque2002}.
Several other systems inherently involve non-conservative terms. For instance, the \textit{Baer–Nunziato model} for compressible two-phase flows includes non-conservative interfacial interaction terms, which are essential for modeling phase exchange dynamics~\cite{Baer1986}. 

In \textit{nonlinear solid mechanics}, especially for materials exhibiting history-dependent or path-dependent behavior, the governing equations frequently appear in non-conservative form~\cite{Godunov2003}.
Although the ideal magneto-hydrodynamics (MHD) equations can be expressed conservatively, the divergence-free condition on the magnetic field ($\nabla \cdot \mathbf{B} = 0$) poses significant numerical challenges. Many schemes introduce non-conservative correction terms to maintain this constraint~\cite{Kemm2002}. Similarly, \textit{advanced traffic flow models}, particularly those accounting for non-local interactions or anticipatory driver behavior, incorporate non-conservative terms to accurately capture emergent dynamics~\cite{aw2000resurrection}.
The \textit{Richards equation}, which models unsaturated groundwater flow in porous media, combines mass conservation with a non-linear form of Darcy’s law and cannot be written in conservative form~\cite{celia1990general}. In the case of \textit{non-Newtonian fluids}, the shear stress terms in the momentum equations are often non-conservative, especially when complex rheological models are involved~\cite{tanner2013rheology}. Similarly, the \textit{Coriolis force} arising in rotating systems generally cannot be cast in a conservative form~\cite{goldstein2002classical}.

To ensure that non-conservative numerical schemes predict the correct shock speed, several strategies have been developed, each addressing different aspects of shock dynamics and entropy consistency. One widely used approach is the class of \textit{path-conservative methods}~\cite{castro2006well},~\cite{dal1995definition}, which generalizes the notion of fluxes by integrating along paths in state space. These methods ensure a consistent treatment of non-conservative products while preserving key conservation properties of the system.
Another effective strategy involves introducing artificial viscosity terms in a manner that respects the entropy structure of the continuous equations. When combined with suitable entropy-stable discretizations, this approach helps enforce the correct selection of physically admissible shock solutions~\cite{fjordholm2012accurate}.
Augmented Riemann solvers provide yet another powerful tool: by explicitly incorporating non-conservative terms into the wave decomposition, they improve the resolution of discontinuities and shock speeds~\cite{berthon2007augmented}. Similarly, \textit{well-balanced schemes} are designed to exactly preserve steady-state solutions, thereby reducing spurious oscillations near shocks and aiding in the accurate prediction of shock propagation~\cite{greenberg1996well}.

Additional techniques to improve the accuracy of numerical schemes include \textit{entropy fixes} and \textit{flux modifications}, which are designed to enforce entropy conditions in discrete settings~\cite{harten1983upstream}. Another important approach is \textit{adaptive mesh refinement (AMR)}, which locally enhances resolution near shocks, thereby improving the accuracy of non-conservative schemes in complex flow regimes~\cite{berger1989local}. Collectively, these strategies help non-conservative formulations achieve reliable shock predictions while addressing the challenges posed by discontinuities and entropy violations. Furthermore, spurious oscillations can be partially suppressed by applying a non-conservative correction to the total energy through a coupled evolution equation for pressure, as demonstrated in~\cite{fedkiw2002general}.

The procedures mentioned above allow non-conservative schemes to handle problems involving shocks and discontinuities. Although their implementation is generally straightforward, stabilizing the solver is slightly difficult because of a lack of a popular solver to handle a variety of flows. Among these, artificial viscosity-based entropy fix methods are relatively straightforward to implement; however, they require extensive tuning to determine the appropriate artificial viscosity coefficient for a given problem. Most existing solvers rely on suboptimal artificial viscosity values. These algorithms can be fine-tuned using deep learning techniques by training on multiple forward problems~\cite{ray2021data},~\cite{bois2023optimal}.

A more direct and effective approach is to use Physics-Informed Neural Networks (PINNs) with adaptive viscosity~\cite{coutinho2023physics},~\cite{neelan2024physics}, where the PINNs framework automatically learns the optimal viscosity required to stabilize the solution. In this work, we adopt this strategy and investigate whether it can effectively handle shocks in both conservative and non-conservative formulations of partial differential equations. Specifically, we focus on the Burgers' equation, and both steady and unsteady Euler equations.

The shock speed estimation is a critical step in high-speed solvers, which is estimated using the Rankine-Hugoniot (RH) jump condition. The RH condition for the conservative equation is:

\begin{equation} \label{eq:shock_speed}
	s = \frac{f(u_R) - f(u_L)}{u_R - u_L}
\end{equation}
Equation~\eqref{eq:shock_speed}  is derived and defined for the conservative form, so the conservative form is commonly used in computational fluid dynamics. When we use the non-conservative form on a problem having shocks and solve it using numerical methods, it gives the wrong shock speed. 

\section{What is Conservation?}

Conservation refers to the preservation of a physical quantity under system evolution. In continuum mechanics, conservation laws arise from integral balance principles \cite{LeVeque2002}:

\begin{equation}
	\frac{d}{dt}\int_{\Omega} U \, d\Omega
	=
	- \int_{\partial \Omega} \mathbf{F}(U)\cdot \mathbf{n} \, dS
	+
	\int_{\Omega} S \, d\Omega,
	\label{eq:integral_form}
\end{equation}

which states that the rate of change of a quantity within a control volume equals the net flux across its boundary plus volumetric sources. Using the divergence theorem, this leads to the local conservative form:

\begin{equation}
	\frac{\partial U}{\partial t}
	+
	\nabla \cdot \mathbf{F}(U)
	=
	S.
	\label{eq:differential_form}
\end{equation}

While this formulation expresses conservation at the continuous level, the notion of conservativeness extends beyond the governing equations and must also be considered at numerical and physical levels \cite{toro2009riemann}.

\subsection{Types of Conservativeness}

\subsubsection{Mathematical Conservativeness}

A system is mathematically conservative if it can be expressed in divergence (flux) form as shown in Eq. \eqref{eq:differential_form}. This ensures:

\begin{itemize}
	\item Consistency with integral balance laws,
	\item Existence of weak solutions \cite{lax1954weak},
	\item Correct discontinuity propagation through Rankine--Hugoniot conditions.
\end{itemize}

Mathematical conservativeness is a property of the governing equations. However, it does not guarantee that a numerical approximation will converge to the correct weak solution.

\subsubsection{Numerical Conservativeness}

A numerical method is conservative if it preserves the discrete integral balance:

\begin{equation}
	\frac{d}{dt} \sum_i U_i \Delta V_i
	=
	\text{boundary fluxes}
	+
	\text{discrete sources}.
\end{equation}

This is typically achieved through flux-difference formulations such as finite volume methods, where interface fluxes are shared between adjacent cells. As noted by \cite{LeVeque2002}, numerical conservativeness ensures:

\begin{itemize}
	\item Exact conservation at the discrete level (up to machine precision),
	\item Correct shock speeds (Telescoping property),
	\item Convergence toward admissible weak solutions (Lax-Wendroff Theorem).
\end{itemize}

Without numerical conservation, even stable and consistent schemes may yield incorrect discontinuity locations and accumulated global errors.

\subsubsection{Physical Conservativeness}

Physical conservativeness refers to the preservation of fundamental physical principles beyond simple mass/momentum balance. This includes:

\begin{itemize}
	\item \textbf{Positivity:} Ensuring density ($ \rho > 0 $), pressure, or water depth remain non-negative.
	\item \textbf{Entropy consistency:} Ensuring the solution satisfies the Second Law of Thermodynamics to avoid non-physical expansion shocks \cite{harten1983high}.
	\item \textbf{Well-balanced property:} The preservation of equilibrium states (e.g., hydrostatic balance in shallow water equations).
\end{itemize}

A scheme may satisfy mathematical and numerical conservation while violating physical constraints, leading to spurious oscillations or simulation crashes.

\subsubsection{Why Conservativeness is Necessary}

Conservativeness is required for three primary reasons:

\begin{enumerate}
	\item \textbf{Correct discontinuity propagation:} Non-conservative discretizations may converge to incorrect shock speeds \cite{hou1994non}.
	\item \textbf{Long-time stability:} Small conservation errors accumulate and cause drift in atmospheric and climate simulations.
	\item \textbf{Physical credibility:} Violation of physical constraints results in nonphysical states and unreliable predictions.
\end{enumerate}

\subsubsection{Hierarchy of Conservativeness}

The three notions form a hierarchy:

\begin{equation}
	\text{Physical} \supset \text{Numerical} \supset \text{Mathematical}.
\end{equation}

Mathematical conservativeness defines the correct continuous model, numerical conservativeness ensures correct discrete representation, and physical conservativeness guarantees conservation and long-term stability. Table~\ref{tab:conservativeness_hierarchy} shows the hierarchy of conservativeness in computational modeling. There are few problems where we no need to worry about these hierarchy when solving but some problems different levels of considerations is required. We will study those in upcoming sections. 

\begin{table}[h!]
	\centering
	\caption{Hierarchy of conservativeness in computational modeling.}
	\vspace{0.2cm}
	\begin{tabular}{llll}
		\toprule
		\textbf{Level} & \textbf{What is Conserved} & \textbf{Enforced By} & \textbf{Failure Leads To} \\
		\midrule
		Mathematical & PDE invariants & Divergence form & Incorrect weak solution \\
		Numerical & Discrete integrals & Flux-consistent schemes & Wrong shock speed \\
		Physical & Fundamental principles & Constraints & Instability \\
		\bottomrule
	\end{tabular}
	\label{tab:conservativeness_hierarchy}
\end{table}

\subsection{Illustration: Non-conservative form leads to wrong shock speed}
We shall try to understand it through one example.
Consider the inviscid Burgers' equation:
\begin{equation}
	\frac{\partial u}{\partial t} + \frac{\partial}{\partial x} \left( \frac{u^2}{2} \right) = 0,
\end{equation}
which is a scalar conservation law with flux function:
\begin{equation}
	f(u) = \frac{u^2}{2}.
\end{equation}

Lets assume we have following Riemann initial condition
\[
u_L = 2 \quad \text{(left state)}, \qquad u_R = 0 \quad \text{(right state)}.
\]

The Rankine--Hugoniot (RH) condition gives the shock speed \( s \) as:
\begin{equation}
	s = \frac{f(u_R) - f(u_L)}{u_R - u_L}
	= \frac{\frac{u_R^2}{2} - \frac{u_L^2}{2}}{u_R - u_L}
	= \frac{1}{2}(u_R + u_L).
\end{equation}

Substituting the values:
\begin{equation}
	s = \frac{1}{2}(0 + 2) = 1.
\end{equation}

{Therefore, the shock propagates to the right with speed \( s = 1 \)}.
In contrast, if one writes the Burgers' equation in non-conservative form:
\begin{equation}
	\frac{\partial u}{\partial t} + u \frac{\partial u}{\partial x} = 0,
\end{equation}
The expression becomes ill-defined at the discontinuity due to the ambiguity in interpreting the product \(u\frac {\partial u}{\partial x} \). Thus, the non-conservative form fails to correctly capture the shock speed.

\subsection{Illustration: The term in the conservative equation should be physically conserved for correct wave speed} 
Another requirement to get the accurate shock speed is that the quantity in the conservative form of PDE should be physically conservative; otherwise, it may lead to the wrong shock speed. This is the illustration based on the example given in Toros' books on \textit{Riemann Solvers and Numerical Methods for Fluid Dynamics: A Practical Introduction }~\cite{toro2009riemann}. 
The 1-D shallow water equation is:

\begin{equation}\label{eq:sh1}
	\begin{bmatrix} \phi \\ \phi u   \end{bmatrix}_t
	=
	-
	\begin{bmatrix}
		\phi u  \\
		\phi u^2 +\frac{1}{2}\phi^2
	\end{bmatrix}_x
\end{equation}

\begin{equation}\label{eq:sh2}
	\begin{bmatrix} \phi \\  u   \end{bmatrix}_t
	=
	-
	\begin{bmatrix}
		\phi u  \\
		\frac{1}{2}u^2 +\phi
	\end{bmatrix}_x
\end{equation}
Equation.~\eqref{eq:sh1} and \eqref{eq:sh2} are mathematically the same and conservative but they lead to different shock speed.
Both the equation are in the divergence form but they give different shock speed. That is because the conserved quantity present in the second row of \eqref{eq:sh2} is not physically conservative.

\paragraph{Wave speed estimation using the second row of \eqref{eq:sh1}:}

Using the Rankine-Hugoniot condition on the second row, we get
\begin{equation}
	s \left[\phi u\right] = \left[\phi u^2 + \frac{1}{2} \phi^2 \right],
\end{equation}
where the jump \([\cdot]\) denotes the difference across the shock:
\[
[\cdot] = (\cdot)_r - (\cdot)_l.
\]
Expanding, this gives
\begin{equation}
	s (\phi_r u_r - \phi_l u_l) = \left(\phi_r u_r^2 + \frac{1}{2} \phi_r^2\right) - \left(\phi_l u_l^2 + \frac{1}{2} \phi_l^2\right).
\end{equation}
Solving for the shock speed \(s\), we get
\begin{equation}
		s = \frac{\phi_r u_r^2 + \frac{1}{2} \phi_r^2 - \phi_l u_l^2 - \frac{1}{2} \phi_l^2}{\phi_r u_r - \phi_l u_l}.
\end{equation}
Let's try to estimate this using a numerical example.
Given the left and right states
\[
\phi_l = 2.0, \quad u_l = 3.0, \quad \phi_r = 1.0, \quad u_r = 1.0,
\]
The wave speed is
\[
s_1 = 3.7.
\]

\paragraph{Wave speed estimation using the second row of \eqref{eq:sh2}:}

Using the Rankine-Hugoniot condition on the second row, we have
\begin{equation}
	s [u] = \left[ \frac{1}{2} u^2 + \phi \right],
\end{equation}
where the jump \([\cdot]\) denotes the difference across the shock:
\[
[\cdot] = (\cdot)_r - (\cdot)_l.
\]
Expanding, this gives
\begin{equation}
	s (u_r - u_l) = \left( \frac{1}{2} u_r^2 + \phi_r \right) - \left( \frac{1}{2} u_l^2 + \phi_l \right).
\end{equation}
Solving for the shock speed \(s\), we get
\begin{equation}
		s = \frac{\frac{1}{2} u_r^2 + \phi_r - \frac{1}{2} u_l^2 - \phi_l}{u_r - u_l}.
\end{equation}

Both the form can give the same shock speed only if $\phi_r=\phi_l$.
Let's try to estimate this using a numerical example.
Given the left and right states
\[
\phi_l = 2.0, \quad u_l = 3.0, \quad \phi_r = 1.0, \quad u_r = 1.0,
\]
and hence
\[
s_2  = 2.5.
\]  

Though \eqref{eq:sh1} and \eqref{eq:sh2} both are in divergence form and valid shallow water equations, they gave the same wave speed for the first equation but a different wave speed for the second equation. When we write the equation in divergence form, it is good to ensure that the conserved variable in the form is a physically conserved quantity to estimate the correct wave speed. In the above example, $u$ is not a conserved variable, but $\phi \times u$ is a conserved variable, so \eqref{eq:sh1} gave the correct wave speed.      
Please note that, so far, we haven't introduced any discretization. So this issue arose before we introduced the numerical discretization. In addition to this, both forms are mathematically equivalent. So the possible limitation is that the wave speed formula is applicable only to conservative discretization. We get $ s(\mathbf{U}_R - \mathbf{U}_L) = \mathbf{F}(\mathbf{U}_R) - \mathbf{F}(\mathbf{U}_L)$ over the control volume only when we use conservative discretization.

\section{Role of conservation on Discretization Framework}

To clearly distinguish the role of mathematical and numerical conservativeness, we briefly describe the discretizations employed in this work.

\subsection{Conservative Flux-Difference Formulation}

For a conservation law written in divergence form,
\begin{equation}
	\frac{\partial U}{\partial t} + \frac{\partial F(U)}{\partial x} = 0,
\end{equation}
a finite-volume discretization yields
\begin{equation}
	\frac{d}{dt} U_i
	=
	-\frac{1}{\Delta x}
	\left(
	F_{i+1/2} - F_{i-1/2}
	\right),
\end{equation}
where $F_{i+1/2}$ denotes a consistent numerical flux shared between adjacent cells.

This formulation ensures exact discrete conservation:
\begin{equation}
	\frac{d}{dt} \sum_i U_i \Delta x
	=
	F_{1/2} - F_{N+1/2}.
\end{equation}

Such schemes converge to weak solutions satisfying the Rankine--Hugoniot condition and therefore produce correct shock speeds under mesh refinement.
\subsection{Non-Conservative Formulation}

Alternatively, the system may be written in quasi-linear form,
\begin{equation}
	\frac{\partial U}{\partial t}
	+
	A(U)\frac{\partial U}{\partial x}
	=
	0,
\end{equation}
which does not explicitly express flux differences.

A standard discretization reads
\begin{equation}
	\frac{d}{dt} U_i
	=
	- A(U_i)
	\frac{U_{i+1}-U_{i-1}}{2\Delta x}.
\end{equation}

Although consistent and potentially stable, this formulation does not preserve a discrete flux balance. Consequently, it may converge to incorrect weak solutions and yield erroneous shock speeds. We will study the effect of this on PINNs and numerical methods in upcoming sections.

\subsection{Path-Conservative and Viscosity-Corrected Form}

To recover correct discontinuity propagation within a non-conservative framework, we incorporate either:

\begin{enumerate}
	\item A path-integral treatment of non-conservative products, or
	\item An adaptive artificial viscosity consistent with jump conditions.
\end{enumerate}

In the path-conservative formulation, interface contributions are defined as
\begin{equation}
	\int_0^1 A(\Phi(s)) \frac{d\Phi}{ds} \, ds,
\end{equation}
where $\Phi(s)$ is a path connecting left and right states.
This construction restores consistency with generalized Rankine--Hugoniot conditions.
Alternatively, artificial viscosity terms of the form
\begin{equation}
	\frac{\partial}{\partial x}
	\left(
	\mu(U) \frac{\partial U}{\partial x}
	\right)
\end{equation}
are introduced, where $\mu(U)$ is adaptively designed to activate near discontinuities.

\subsection{Test Case 1: Mathematical Conservativeness via Burgers' Equation}
\label{sec:test1}
To evaluate mathematical conservativeness and weak-solution behavior, we consider the inviscid Burgers' equation in strictly conservative form:
\begin{equation}
	\frac{\partial u}{\partial t} + \frac{\partial}{\partial x} \left( \frac{u^2}{2} \right) = 0, \quad x \in [-1,1].
	\label{eq:burgers}
\end{equation}
We prescribe a Riemann initial condition with $u_L=1$ ($x<0$) and $u_R=0$ ($x \ge 0$). This generates a shock wave with a propagation speed determined by the Rankine--Hugoniot condition, $s = \Delta f / \Delta u = 0.5$, yielding the exact shock location $x_s(t) = 0.5t$.
To demonstrate the impact of conservativeness, we compare two distinct discretizations:

\paragraph{1. Conservative Discretization}
We employ a standard first-order upwind scheme based on the flux formulation. The update formula preserves the cell averages:
\begin{equation}
	u_i^{n+1} = u_i^n - \frac{\Delta t}{\Delta x} \left( F_{i+1/2} - F_{i-1/2} \right),
\end{equation}
where the numerical flux is given by the Godunov flux for the scalar Burgers equation:
\begin{equation}
	F_{i+1/2} = \max \left( f(u_i^n), f(u_{i+1}^n) \right) \quad \text{if } u_i > u_{i+1}.
\end{equation}
This formulation ensures that $\sum u_i$ is conserved (telescoping sum property) and converges to the correct shock speed $s=0.5$.

\paragraph{2. Non-Conservative Discretization}
Alternatively, we discretize the quasi-linear form $u_t + u u_x = 0$ using non-conservative upwinding:
\begin{equation}
	u_i^{n+1} = u_i^n - \frac{\Delta t}{\Delta x} u_i^n \left( u_i^n - u_{i-1}^n \right), \quad \text{for } u_i > 0.
\end{equation}
Although this scheme is consistent ($O(\Delta x)$ truncation error) and stable under CFL conditions, it does not satisfy the discrete integral balance. As shown by Hou and LeFloch (1994), this form converges to a non-physical weak solution with an incorrect shock speed of $s=0$.

\subsubsection*{Numerical Setup}
\begin{figure}
	\centering
	\begin{subfigure}[b]{0.49\textwidth}
		\includegraphics[width=\textwidth]{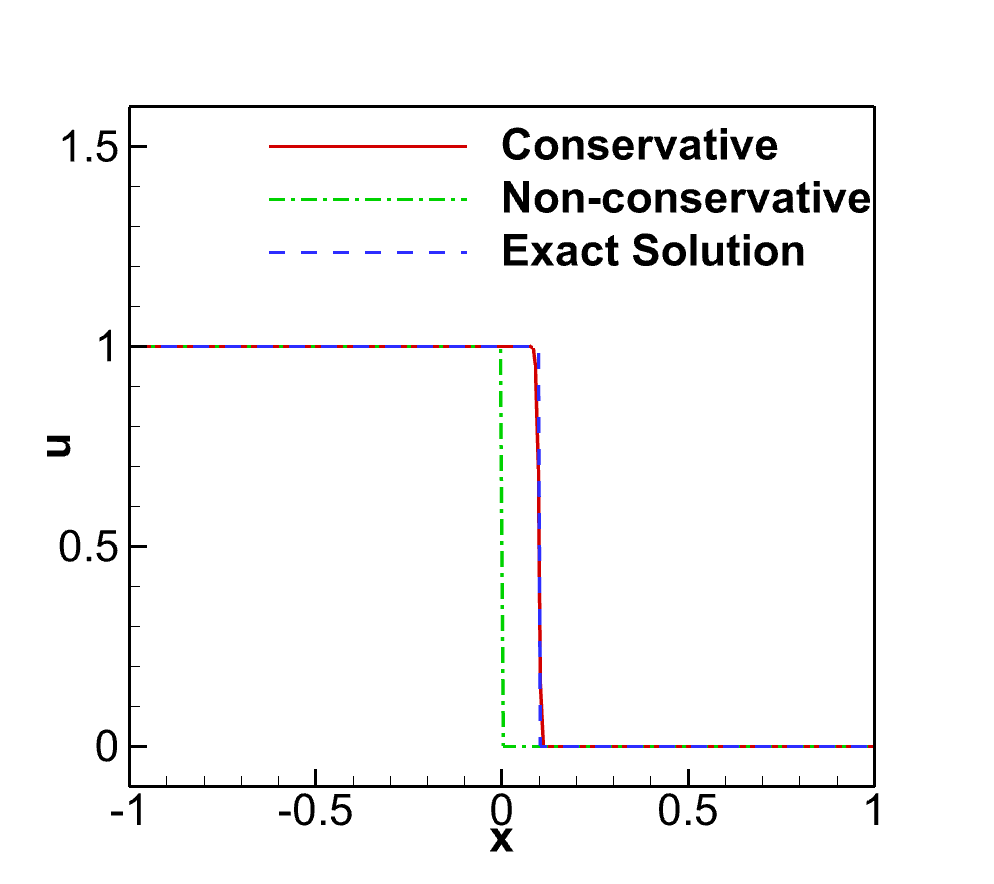}
		\caption{Burgers equation }
		\label{fig:burg_non1}
	\end{subfigure}
	\begin{subfigure}[b]{0.49\textwidth}
		\includegraphics[width=\textwidth]{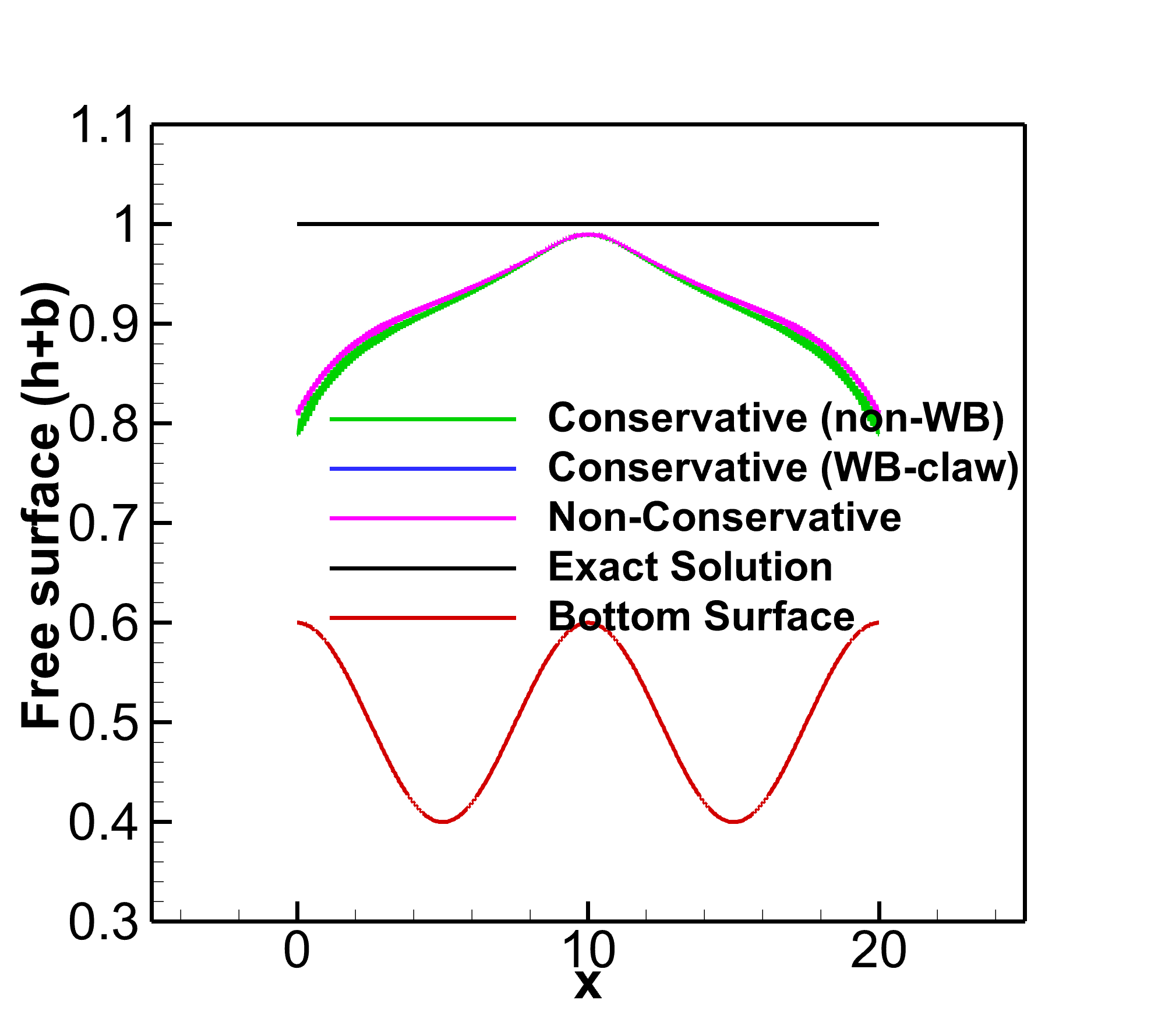}
		\caption{Shallow water equation}
		\label{fig:SWL_p}
	\end{subfigure}
	\caption{Solution using conservative and non-conservative scheme }
	\label{fig:bu1}
\end{figure}

The domain is discretized using a uniform grid with $N_x = 400$. Time integration is performed using an explicit scheme with $\text{CFL} = 0.4$ up to $t_{\text{final}} = 0.2$ is shown in figure~\ref{fig:burg_non1}. In this figure we can observe that non-conservative scheme unable to resolve the correct shock speed but conservative scheme able locate exact shock speed.

\subsubsection*{Role in Assessing Conservativeness}

This test case isolates mathematical and numerical conservativeness as follows:

\begin{itemize}
	\item The governing PDE is mathematically conservative by construction.
	\item A conservative flux-difference discretization preserves discrete integral balance and converges to the correct weak solution.
	\item A non-conservative discretization, despite consistency and stability, may violate discrete flux balance and yield incorrect shock speeds.
\end{itemize}

\subsection{Test Case 2: Unified Assessment of Conservativeness Using the Shallow Water Equations}
\label{sec:swl}
To simultaneously illustrate mathematical, numerical, and physical conservativeness, we consider the one-dimensional shallow water equations (SWE) with bottom topography:

\begin{align}
	\frac{\partial h}{\partial t} + \frac{\partial (hu)}{\partial x} &= 0, \\
	\frac{\partial (hu)}{\partial t}
	+ \frac{\partial}{\partial x}
	\left( hu^2 + \frac{1}{2} g h^2 \right)
	&= - g h \frac{\partial b}{\partial x}.
\end{align}

Here, $h$ denotes water depth, $u$ velocity, $b(x)$ bottom elevation, and $g$ gravitational acceleration.

\subsubsection*{Mathematical Conservativeness}

The SWE are mathematically conservative in the variables $(h, hu)$ when written in divergence form. The system satisfies integral balance laws and admits physically relevant steady solutions. In particular, the lake-at-rest equilibrium:

\begin{equation}
	u = 0, \qquad h + b = \text{constant},
\end{equation}

represents a non-trivial steady state resulting from exact balance between flux gradients and the source term.

\subsubsection*{Computational Setup}

The computational domain is
\[
x \in [0,L], \quad L = 20,
\]
discretized using $N_x = 400$ uniform grid points with spacing $\Delta x = L/(N_x-1)$.

The bottom profile is prescribed as:
\begin{equation}
	b(x) = b_0 + A_b \cos\left(\frac{2\pi x}{L_b}\right),
\end{equation}
with parameters $b_0=0.5$, $A_b=0.1$, and $L_b=10$.

Initial conditions are:
\begin{align}
	h(x,0) &= H_0 - b(x) + A_p \exp\left(-\frac{(x-x_c)^2}{2\sigma^2}\right), \\
	u(x,0) &= 0,
\end{align}
where $H_0=1.0$, $A_p=0.0$, $x_c=L/2$, and $\sigma=0.6$.

Time integration uses a second-order Runge--Kutta scheme with time step determined by the CFL condition:
$
	\Delta t = \text{CFL} \frac{\Delta x}{\sqrt{gH_0}},
	\quad \text{CFL}=0.4,
$
and final time $T_{\text{end}}=5$.

\subsubsection{Discretizations Used in the SWE Test Case}
\label{sec:swl1}
To isolate the role of different levels of conservativeness, we employ multiple spatial discretizations of the shallow water equations.

\paragraph{1. Quasi-Linear (Non-Conservative) Finite Difference Formulation}

The governing equations are rewritten in primitive-variable form:

\begin{align}
	h_t &= -u h_x - h u_x, \\
	u_t &= -u u_x - g h_x - g b_x.
\end{align}

Spatial derivatives are approximated using second-order centered finite differences:

\begin{equation}
	q_x \approx \frac{q_{i+1} - q_{i-1}}{2\Delta x}.
\end{equation}

Although this formulation is consistent and formally second-order accurate, it does not preserve a discrete flux balance in $(h, hu)$. Consequently:

\begin{itemize}
	\item Total mass is not preserved exactly,
	\item Total momentum exhibits drift,
	\item Equilibrium states are not maintained.
\end{itemize}

This discretization violates numerical conservativeness.

\paragraph{2. Conservative Finite-Volume Formulation}

The SWE are written in conservative vector form:

\begin{equation}
	\frac{\partial \mathbf{U}}{\partial t}
	+
	\frac{\partial \mathbf{F}(\mathbf{U})}{\partial x}
	=
	\mathbf{S}(\mathbf{U}),
\end{equation}

with
\[
\mathbf{U} = 
\begin{pmatrix}
	h \\ hu
\end{pmatrix},
\quad
\mathbf{F}(\mathbf{U}) =
\begin{pmatrix}
	hu \\
	hu^2 + \frac{1}{2}gh^2
\end{pmatrix},
\quad
\mathbf{S} =
\begin{pmatrix}
	0 \\
	- g h b_x
\end{pmatrix}.
\]

A semi-discrete finite-volume formulation reads:

\begin{equation}
	\frac{d}{dt}\mathbf{U}_i
	=
	-\frac{1}{\Delta x}
	\left(
	\mathbf{F}_{i+1/2}
	-
	\mathbf{F}_{i-1/2}
	\right)
	+
	\mathbf{S}_i.
\end{equation}

This formulation guarantees exact discrete conservation of mass:

\begin{equation}
	\frac{d}{dt} \sum_i h_i \Delta x = 0.
\end{equation}

However, unless the source term is discretized consistently with the flux gradient, equilibrium states may still be destroyed.

\paragraph{3. Conservative but Non-Well-Balanced Form}

In many standard implementations, the source term is discretized independently of the flux term:

\begin{equation}
	S_i = - g h_i b_x.
\end{equation}

Although global conservation is preserved, the discrete flux gradient does not exactly balance the source term at steady state. As a result:

\begin{itemize}
	\item Spurious motion develops,
	\item The lake-at-rest equilibrium is not preserved,
	\item Free-surface distortion appears.
\end{itemize}

This demonstrates that numerical conservativeness alone does not ensure physical conservativeness.

\paragraph{4. Well-Balanced Formulation}

A well-balanced scheme is constructed such that the discrete flux gradient and source term cancel exactly for the equilibrium state:

\[
u = 0, \quad h + b = \text{constant}.
\]

This typically requires hydrostatic reconstruction or modified flux evaluation ensuring that:

\begin{equation}
	\left( \frac{1}{2} g h^2 \right)_x + g h b_x = 0
\end{equation}

is satisfied discretely.

Only under this condition are:

\begin{itemize}
	\item Mathematical conservativeness,
	\item Numerical conservativeness,
	\item Physical equilibrium preservation
\end{itemize}

simultaneously satisfied. Figure~\ref{fig:SWL_p} shows the free surface elevation of conservative scheme with and without well balanced treatment, non-conservative scheme. From this it is clear that conservative schemes without well balanced enforcement unable to produce the correct free surface elevation.     
\subsubsection*{Interpretation}

This single test case therefore illustrates:

\begin{itemize}
	\item The governing equations are mathematically conservative.
	\item Non-conservative discretizations violate numerical conservativeness.
	\item Conservative discretizations may still violate physical equilibrium.
	\item Well-balanced schemes are required to ensure physical conservativeness.
\end{itemize}
Conservativeness in computational fluid dynamics is inherently hierarchical. Mathematical conservation defines the governing equations through divergence-form partial differential equations, ensuring integral properties hold in the continuous limit. Numerical conservation translates this to the discrete level, enforcing a rigorous flux balance across cell interfaces to correctly capture wave speeds and jump conditions. However, for many systems, these are necessary but insufficient.

Furthermore, a carefully balanced discretization is crucial to preserve physically meaningful equilibrium states, such as the lake at rest condition in the shallow water equations. Without physics-aware schemes that also enforce shock-consistent entropy for compressible flows, numerical methods may formally conserve quantities but still produce non-physical solutions. Ultimately, ensuring physical fidelity requires targeted enforcement at all three levels of this hierarchy.

\subsection{Can Non-Conservative Schemes Be Used for Shock Problems?}

In general, conservative schemes are necessary for hyperbolic conservation laws containing shocks, since they satisfy the Rankine--Hugoniot jump conditions and converge to the physically correct weak solution \cite{leveque1992numerical}. The Lax--Wendroff theorem \cite{lax1960systems} establishes that any consistent, conservative scheme converges (if it converges) to a weak solution of the conservation law. Consequently, non-conservative discretizations are often considered unsuitable for shock-dominated flows.

Nevertheless, non-conservative formulations are not inherently invalid and can be advantageous in several situations. Many physical models are naturally expressed in primitive variables rather than conserved variables, including low-Mach-number flows, multiphase systems, turbulence closures, and relativistic hydrodynamics. Conservative formulations may also introduce numerical stiffness or loss of accuracy due to cancellation errors, particularly when kinetic energy dominates internal energy. Furthermore, certain systems (e.g., Baer-Nunziato-type multiphase models \cite{Baer1986}) contain non-conservative products that cannot be written purely in divergence form. Non-conservative formulations may therefore offer improved conditioning and modeling flexibility, especially when coupled with physics-informed machine learning approaches.
The primary challenge is that non-conservative schemes may converge to incorrect shock solutions unless additional mechanisms enforce the proper jump conditions.

\paragraph{Path-Conservative (Path-Integral) Approach}

A rigorous framework for non-conservative hyperbolic systems was developed by Dal Maso, LeFloch, and Murat \cite{dal1995definition} through the definition of weak solutions using path integrals in state space. The key idea is to replace the ill-defined non-conservative product across discontinuities with an integral taken along a path connecting the left and right states. Numerically, path-conservative schemes \cite{pares2006numerical} compute interface contributions as
\begin{equation}
	\int_0^1 \mathbf{A}(\boldsymbol{\Phi}(s)) \, \frac{d\boldsymbol{\Phi}}{ds} \, ds,
\end{equation}
where $\boldsymbol{\Phi}(s)$ defines a path in state space between the two states. This formulation generalizes conservative finite-volume methods and recovers the correct Rankine--Hugoniot conditions. Such schemes are widely used for multiphase and non-equilibrium flows, although the choice of path can influence the selected weak solution.

\paragraph{Artificial Viscosity}

Artificial viscosity \cite{von1950method} provides an alternative stabilization mechanism by introducing controlled dissipation near steep gradients. This regularizes discontinuities into thin layers and enforces entropy conditions. Modern artificial viscosity methods include adaptive viscosity based on shock sensors \cite{persson2006sub}, matrix viscosity aligned with characteristic fields, and data-driven or learned viscosity models. While effective for stabilization, excessive dissipation may smear shocks and reduce accuracy.

\paragraph{Hybrid and Entropy-Stable Strategies}

We can combine non-conservative formulations with entropy-stable dissipation \cite{tadmor2003entropy, castro2013entropy}, nonlinear limiters, path-consistent interface treatments, and adaptive viscosity mechanisms. These approaches aim to retain the flexibility of primitive-variable formulations while ensuring physically correct shock capturing.
Non-conservative schemes can be applied to problems with shocks provided that additional mechanisms enforce correct discontinuity behavior. Path-conservative formulations offer a mathematically rigorous framework, whereas artificial viscosity supplies practical stabilization. Such methods are particularly relevant for complex multiphysics systems and emerging physics-informed computational approaches where conservative formulations may be restrictive.

\subsection{Test Case 3: Conservative vs Non-Conservative Formulations for the Sod Shock Tube}

To investigate the behavior of non-conservative formulations in the presence of shocks and the effect of parabolic regularization, we consider the one-dimensional Sod shock tube governed by the Euler equations. The computational domain is $x \in [0,2]$ with an initial discontinuity at $x_0 = 1$. The initial states are
\begin{equation}
	(\rho,u,p) =
	\begin{cases}
		(1.0,\, 0,\, 1.0), & x < x_0, \\
		(0.125,\, 0,\, 0.1), & x > x_0,
	\end{cases}
\end{equation}
with ratio of specific heats $\gamma = 1.4$.
The solution is evolved until $T = 0.5$ using two different numerical approaches.

\paragraph{Conservative Finite-Volume Method}

The Euler equations in conservative form,
\begin{equation}
	\mathbf{U}_t + \mathbf{F}(\mathbf{U})_x = 0,
\end{equation}
are discretized using a first-order finite-volume method with the Rusanov (local Lax--Friedrichs) flux:
\begin{equation}
	\mathbf{F}_{i+\frac{1}{2}} =
	\frac{1}{2}\left(\mathbf{F}_L + \mathbf{F}_R\right)
	- \frac{1}{2}\alpha_{i+\frac{1}{2}}(\mathbf{U}_R - \mathbf{U}_L),
\end{equation}
where
$
	\alpha_{i+\frac{1}{2}} = \max(|u_L|+c_L, |u_R|+c_R)
$
is the maximum signal speed. This conservative scheme satisfies the Rankine--Hugoniot jump conditions and produces the correct weak solution.

\paragraph{Non-Conservative Primitive-Variable Formulation}

For comparison, the Euler equations are also solved in primitive variables $(\rho,u,p)$ using a non-conservative formulation augmented with artificial viscosity:
\begin{align}
	\rho_t + u \rho_x + \rho u_x &= \nu \rho_{xx}, \\
	u_t + u u_x + \frac{1}{\rho} p_x &= \nu u_{xx}, \\
	p_t + u p_x + \gamma p u_x &= \nu p_{xx}.
\end{align}

\begin{figure}
	\centering
	\includegraphics[width=0.7\linewidth]{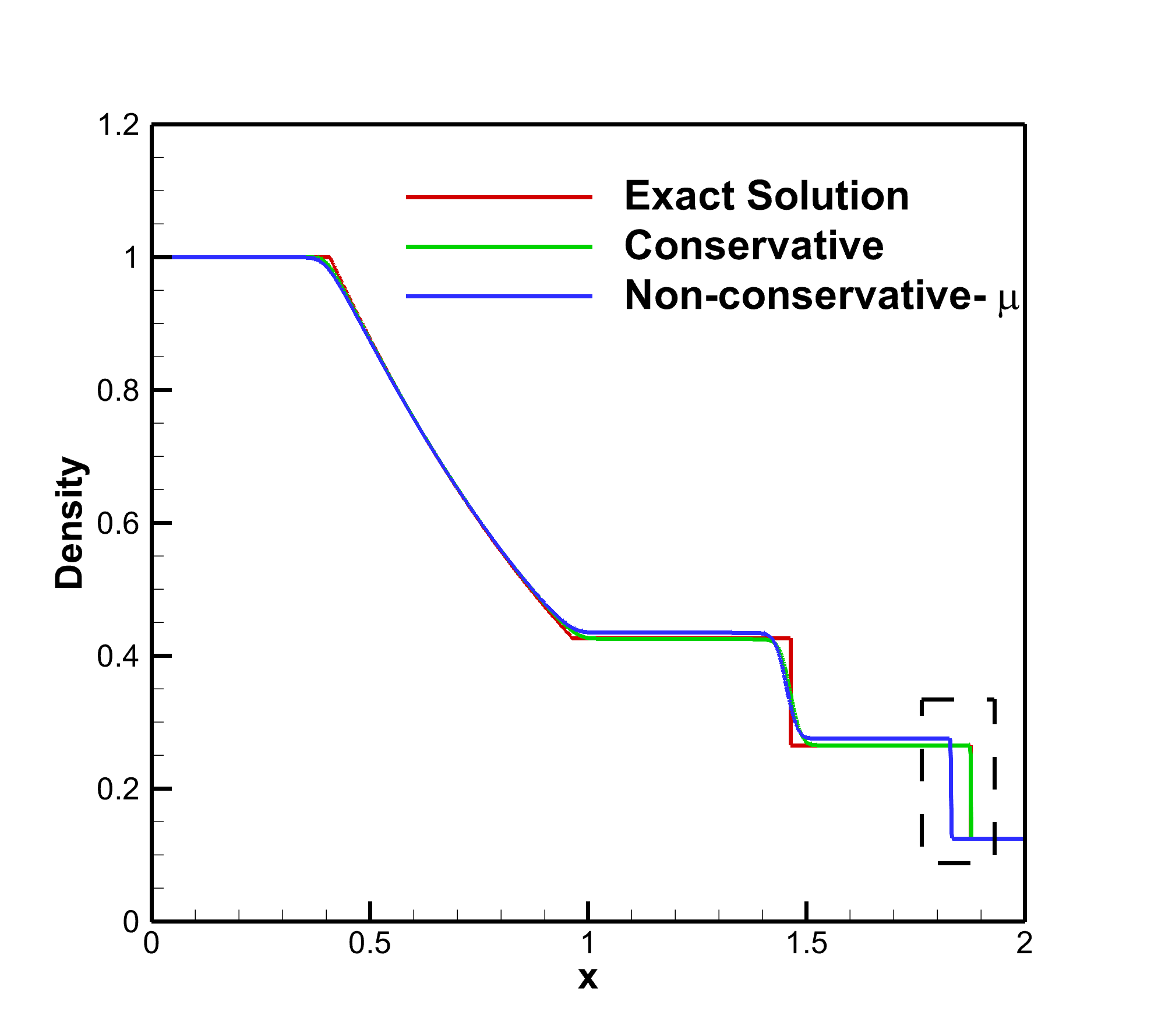}
	\caption{Density of Sod shock tube problem at T = 0.5 s}
	\label{fig:connoncon}
\end{figure}

Spatial derivatives are approximated using central finite differences. The viscosity coefficient is chosen adaptively as
\begin{equation}
	\nu = \frac{1}{2} a_{\max} \Delta x,
\end{equation}
where $a_{\max} = \max(|u| + c)$ is the maximum local wave speed. This diffusion term introduces parabolic regularization analogous to Rusanov-type dissipation. Figure~\ref{fig:connoncon} demonstrates that the non-conservative scheme with viscous dissipation (non-conservative-$\mu$) fails to predict the correct shock speed. As time progresses, the shock position obtained from the non-conservative formulation increasingly deviates from the exact solution, while the numerical solution remains apparently stable. This behavior is noteworthy, as stability alone does not guarantee physical accuracy. Table~\ref{tab:com} summarize the different level of conservation required for the different equations considered in this work.

\begin{table}[htbp]
	\centering
	\caption{Comparison of conservation requirements across different physical systems.}
	\label{tab:conservation_hierarchy}
	\renewcommand{\arraystretch}{1.2} 
	\begin{tabular}{l c c c c}
		\hline
		\textbf{System / Problem} & \textbf{Math. Cons.} & \textbf{Num. Cons.} & \textbf{Well-Balanced} & \textbf{Shock/Entropy} \\
		\hline
		Burgers' (Inviscid) & Required & Required & N/A & Essential \\
		Shallow Water (Static) & Required & Required & Essential & Optional \\
		Euler Equations & Required & Required & Recommended & Essential \\
		\hline
	\end{tabular}
\label{tab:com}
\end{table}

\section{Introduction to PINNs}     
In this section, we have presented the details of the basic PINNs architecture and the adaptive weight and viscosity architecture used in this work.
\subsection{Introduction}
Modern CFD possesses a powerful toolkit with Finite Difference, Finite Element, Finite Volume, Spectral Element, and meshless approaches such as Smoothed Particle Hydrodynamics. However, there are still significant hurdles that are inherently difficult for CFD to tackle. For one, the incorporation of available empirical data is a rich source of physical information and is incorporated into CFD through correlations and models, and not directly. CFD also requires the governing system of equations to be complete, i.e., all boundary conditions must be known to a desired order of accuracy especially for atmospheric simulations. This is a severe limitation to the accuracy that CFD can extend to; for instance, available BC restricts CFD to the Navier-Stokes system, and advanced hydrodynamics cannot be captured through Burnett and super Burnett equations.

Statistical learning, a subfield of statistics focused on establishing a relationship between variables in a dataset and the output, has been modernized as Machine Learning (ML). The key principle of ML is to develop algorithms that learn from data to identify patterns, and make accurate predictions, with the added ability to continuously learn through experience. Rapid advancements in ML have revolutionized CFD research in many fields, offering a powerful data-driven approach. However, purely data-driven models lack the physical consistency and can lead to the learning process going astray. This has been observed by many researchers, and the unphysical results reported are an interesting counterpoint to the applicability of ML in CFD. Although well-trained models are great interpolators, they struggle when tasked with extrapolating beyond the training data.

The inspiration to include the rich theoretical knowledge into the data-driven ML process has resulted in a potent computational framework named Physics-Informed Neural Networks (PINNs). In this class of methods, the network’s loss function, which consists of data loss, is augmented by a PDE loss, which signifies the deviation of the solution from the physical principles cast as a partial differential equation. In this paradigm, efforts to reduce the loss function is essentially efforts to ensure data and physical-fidelity of the solution.

The success of PINNs, specifically for solution to PDEs, can be attributed to their ability to leverage automatic differentiation (autograd or AD), which enables efficient and accurate computation of necessary derivatives to associate the inputs and outputs of the neural network. This feature enables PINNs to directly enforce the physics that are expressed through differential equations.

\subsection{PINNs process}
The entire process of setting up the PINNs for PDEs can be summarized by the following 3 steps,
\begin{itemize}
	
	\item A neural network takes spatial and temporal coordinates as input and outputs the predicted solution to the physical system (e.g., temperature, velocity, displacement). Often, a fully connected feed-forward network is utilized for the network architecture.
	
	\item Automatic differentiation (AD) is a crucial component of PINNs. It allows for the efficient and accurate computation of the derivatives of the neural network's output with respect to its inputs. These derivatives are precisely what's needed to evaluate the terms in the differential equations for the physics loss. 
	
	\item During training, an optimization algorithm (like Adam or L-BFGS) iteratively adjusts the neural network's parameters to minimize the total loss function (data loss + physics loss).
\end{itemize}

Developing a PINNs circuit from ground-up can be a daunting prospect from a coding standpoint. However, this is offset by the availability of standard frameworks such as PyTorch and TensorFlow, which have made the process extremely straightforward. However, despite the promise, PINNs face numerous challenges that need to be addressed. A large portion of these challenges is associated with enormous computational cost, while the next significant challenge arises due to generalizability. 
Computational cost can be tackled by decreasing the depth of the PINNs, which means there are fewer hidden layers and lesser number of trainable parameters. However, this severely restricts the expressive power of the PINNs. A complementary approach is to decompose the computational domain into multiple sub-domains and utilize a separate, low depth PINNs for each sub-domain. This approach lends itself to parallel computing naturally and reduces the cost associated with each PINNs (\cite{dwivedi2019distributed,SHUKLA2021110683}).  Such Distributed PINNs (DPINNs) have been extensively explored in the context of CFD, resulting in several customized PINNs. Two examples of this philosophy are the Conservative PINNs (C-PINNs) \cite{JAGTAP2020113028} and the eXtended PINNs (XPINNs) \cite{CiCP-28-5}, which vary in their decomposition strategies and problem types they can handle.

Conservative PINNs utilize a domain decomposition approach and are particularly well-suited to solve problems governed by strong conservation laws. Considering that the basis of CFD is the conservation of mass, momentum, and energy, these PINNs have shown a very relevant contribution to the present discussion. The critical challenge that is addressed by cPINNs is the enforcement of flux continuity across the interfaces of the decomposed domain, which is explicitly enforced through the strong form of the governing equations. Discrepancies in the enforcement are added to the loss functions and act as a penalty, which directs the solution towards a conservative solution. 
While cPINNs primarily focus on dividing the domain along its spatial dimensions and implementing shallow neural networks in each subdomain to learn the solution within these spatial sub-domains, XPINNs build upon and generalize cPINNs. XPINNs/Distributed PINNs expand the domain decomposition concept to both space and time. This allows them to handle a wider range of PDEs and employ arbitrary decomposition strategies.

Another approach to tackle the computational cost is to reduce the training time of the PINNs. This line of thinking has led to another variation of PINNs called the Variational PINNs (V-PINNs) \cite{kharazmi2019variationalphysicsinformedneuralnetworks}. Unlike the PINNs discussed above, VPINNs deviate from the strong form of the governing equations and instead incorporate the weak form into their loss function. Traditionally, this approach is critical where high-order differentiability of the system is not guaranteed. In the context of PINNs, this approach reduces the load of the auto differentiation calculations and thus accelerates the PINNs computations. Procedurally, VPINNs use test functions and quadrature points instead of a random collocation of points to compute the integrals and consequently, the variational residuals. This replacement of the training dataset by quadrature points has demonstrated a reduction in the training time. Although VPINNs require a mesh to define the integrals and basis functions, Mesh-Free Variational PINNs (MF-VPINNs) have been developed since to address this issue \cite{Berrone_2024}. 

Another variant of PINNs that utilizes the integral form of the conservation laws is called the Integral PINNs (IPINNs)~\cite{rajvanshi2024integral}. This variant has immense utility for problems that include shock discontinuities where the differential form of the governing equation does not hold. 
Finite Basis PINNs (FBPINNs)~\cite{moseley2023finite} methods combine PINNs with concepts from finite basis functions (e.g., Fourier series, wavelets) to improve the representation capacity and efficiency, particularly for oscillatory or multi-scale problems.

Acceleration of PINNs training can also be achieved through Adaptive PINNs (APINNs)~\cite{xiang2022self}, which dynamically adapt various aspects of the PINNs during training. This can involve Adaptive Sampling, which adjusts the distribution of collocation points based on the PDE residual or solution error to focus computational resources on challenging regions. Adaptive Weighting strategy dynamically adjusts the weights of different terms in the loss function (e.g., physics loss, data loss, boundary condition loss) to balance their contributions and improve convergence. Adaptive Activation Functions can change their parameters during training to better fit the solution.

Another line of research close to PINNs is that of Operator Learning methods~\cite{patel2024variationally}. While distinct from standard PINNs, these are closely related and represent a powerful class of "operator learning" methods. Instead of learning a specific solution to a PDE, they learn the operator that maps an input function (e.g., initial condition, boundary condition, source term) to the solution function. Once trained, a PINO/DeepONet can rapidly predict solutions for any new input function without retraining, making them incredibly efficient for many-query problems. In particular, DeepONets introduced in \cite{Lu_2021} are inspired by the Universal Approximation Theorem of Neural Networks and attempt to learn the solution operator. A popular physics-informed variation of this operator called Physics-Informed Neural Operator (PINO) was introduced in \cite{li2024physics}. PINO makes use of Fourier Neural Operator (FNO) \cite{li2020fourier} and combines training data and physics constraints to learn the solution operator for a given family of parametric PDEs.

Physics-Informed Neural Networks (PINNs) present unique challenges when applied to supersonic flows due to the difficulty of computing gradients across shock discontinuities. Nevertheless, they have demonstrated the capability to capture shock structures and discontinuities without relying on traditional numerical solvers~\cite{mao2020physics}. PINNs have also proven effective for modeling multi-phase flows involving shocks, without the need for explicit shock-capturing techniques~\cite{wang2021learning}. To improve robustness in shock-dominated regimes, a hybrid approach combining mesh-aware training and shock-aware viscosity control within the PINNs framework was proposed in~\cite{wassing2024physics}. Furthermore, PINNs have been shown to match the performance of traditional shock-fitting and shock-capturing solvers in predicting both continuous and discontinuous flow features in nozzle geometries, enabling data-free, high-fidelity CFD using neural approaches~\cite{liang2024continuous,KUMAR2026106975}. Recent work also introduced a reinitialization strategy tailored for stiff problems, such as high-Reynolds-number flows, enhancing the convergence and accuracy of PINNs training in such regimes~\cite{lee2025physics}.

\subsection{PINNs to solve PDEs}
In this section, we shall see the procedure to solve PDEs using PINNs. Let's consider the 1D heat equation:
\[
\frac{\partial u}{\partial t} = \alpha \frac{\partial^2 u}{\partial x^2}, \quad x \in [0,1], \, t \in [0,T]
\]
With the initial condition:
\[
u(x, 0) = u_0(x)
\]
and boundary conditions:
\[
u(0,t) = u_L(t), \quad u(1,t) = u_R(t)
\]
As per the universal approximation theorem, neural networks can approximate any continuous function. So,
we can approximate the solution of the heat equation using a neural network:
\[
u(x,t) \approx u_\theta(x,t)
\]
where $\theta$ denotes the parameters (weights and biases) of the neural network. The random weight and bias cannot satisfy the PDEs, so we will get some residue, which can be written as
\[
\mathcal{R}(x,t) = \frac{\partial u_\theta}{\partial t} - \alpha \frac{\partial^2 u_\theta}{\partial x^2}
\]
The physics-based loss (PDE loss) over all sample points can be written as
\[
\mathcal{L}_{\text{PDE}} = \frac{1}{N_f} \sum_{i=1}^{N_f} \left| \mathcal{R}(x_f^i, t_f^i) \right|^2
\]
where $\{(x_f^i, t_f^i)\}_{i=1}^{N_f}$ are collocation points in the interior domain.

\[
\mathcal{L}_{\text{IC}} = \frac{1}{N_0} \sum_{i=1}^{N_0} \left| u_\theta(x_0^i, 0) - u_0(x_0^i) \right|^2
\]
The PDE can give a unique solution only when we apply the initial and boundary conditions.
\[
\mathcal{L}_{\text{BC}} = \frac{1}{N_b} \sum_{i=1}^{N_b} \left( 
\left| u_\theta(0, t_b^i) - u_L(t_b^i) \right|^2 + 
\left| u_\theta(1, t_b^i) - u_R(t_b^i) \right|^2 
\right)
\]

The total loss is a weighted sum of all loss components:
\[
\mathcal{L}_{\text{total}} = \lambda_{\text{PDE}} \mathcal{L}_{\text{PDE}} + 
\lambda_{\text{IC}} \mathcal{L}_{\text{IC}} + 
\lambda_{\text{BC}} \mathcal{L}_{\text{BC}}
\]
where $\lambda_{\text{PDE}}, \lambda_{\text{IC}}, \lambda_{\text{BC}}$ are weights that can be tuned to balance the contributions. 
The network parameters $\theta$ are optimized using gradient-based methods:
\[
\theta^* = \arg \min_\theta \mathcal{L}_{\text{total}}(\theta)
\]

\subsection{Adaptive weight and viscosity PINNs architecture (PINNs-AWV)}
Adaptive weight and viscosity PINNs architecture is a sophisticated neural network architecture designed to handle shocks and discontinuity was introduced in~\cite{neelan2024physics}. Most of the problems involving shocks and discontinuities are modeled using differential equations. When we calculate the derivative across the shocks, the residue of the PDEs has a high value, which will destabilize the solver. Some of the ways to handle this are by reducing the weights at these regions or adding artificial viscosity to the solver to spread the shocks over several grid points to reduce the magnitude of the derivatives across the shocks. This architecture uses both features to stabilize the solver. We have used 1-D Burgers' Equation to explain the mathematical details of the   the PINNs-AWV architecture. The 1D Burgers' Equation is

\begin{equation}
	\frac{\partial u}{\partial t} + u \frac{\partial u}{\partial x} = 0
\end{equation}

To smooth the shock produced by the Burgers equation, we add an artificial viscosity term to the Burgers equation to stabilize the solver, but we don't know the optimal viscosity to stabilize the solver. If we add a large amount, this will reduce the shock resolution property of the scheme. If we add too much or not enough will destabilize the solver~\cite{drozda2023learning}.
In traditional numerical methods, we test the solvers on several test cases and find the optimal viscosity that works. We can also learn the optimal viscosity, which stabilize the solver using the neural network~\cite{ray2021data},~\cite{dubey2021learning}. The Burgers equation with artificial viscosity can be written as:

\[
\frac{\partial u}{\partial t} + u \frac{\partial u}{\partial x} = \nu \frac{\partial^2 u}{\partial x^2}
\]

where $x \in [-1, 1]$, $t \in [0, T]$, and $\nu$ is the artificial viscosity co-efficent.
We approximate the solution as:
\[
u(x,t) \approx u_\theta(x,t)
\]
with neural network parameters $\theta$.
Where $\nu$ is the trainable parameter modeled via a sub-network. Because the Burgers equation does not have viscosity, we should minimize the viscosity. The PINNs will find the optimal viscosity to stabilize the solver without training data. So the viscous loss can be written as 
\[
\mathcal{L}_{\nu} = \nu^2 
\]
The PDE residual becomes:
\[
\mathcal{R}(x,t) = \frac{\partial u_\theta}{\partial t} + u_\theta \frac{\partial u_\theta}{\partial x} - \nu(x,t) \frac{\partial^2 u_\theta}{\partial x^2}
\]
Considering initial and boundary losses

\[
\mathcal{L}_{\text{IC}} = \frac{1}{N_0} \sum_{i=1}^{N_0} \left| u_\theta(x_0^i, 0) - u_0(x_0^i) \right|^2
\]

\[
\mathcal{L}_{\text{BC}} = \frac{1}{N_b} \sum_{i=1}^{N_b} \left( 
|u_\theta(-1, t_b^i) - u_L(t_b^i)|^2 + |u_\theta(1, t_b^i) - u_R(t_b^i)|^2 
\right)
\]
When we use weights $\lambda_{\text{PDE}}$ that adapt based on gradient norms:
\[
\lambda_{PDE} = \frac{1}{\left\|\nabla_\theta \mathcal{L}_i \right\| + \epsilon}
\]
The total loss is
\[
\mathcal{L}_{\text{total}} = \tilde{\lambda}_{\text{PDE}} \mathcal{L}_{\text{PDE}} + 
\tilde{\lambda}_{\text{IC}} \mathcal{L}_{\text{IC}} + 
\tilde{\lambda}_{\text{BC}} \mathcal{L}_{\text{BC}}+ 
\tilde{\lambda}_{\nu} \mathcal{L}_{\nu}
\]

All parameters, including neural network weights $\theta$ and possibly $\nu(x,t)$, are optimized by minimizing the total loss:
\[
\theta^*, \nu^* = \arg \min_{\theta, \nu} \mathcal{L}_{\text{total}}(\theta, \nu)
\]

\begin{figure}
	\centering
	\includegraphics[width=0.8\linewidth]{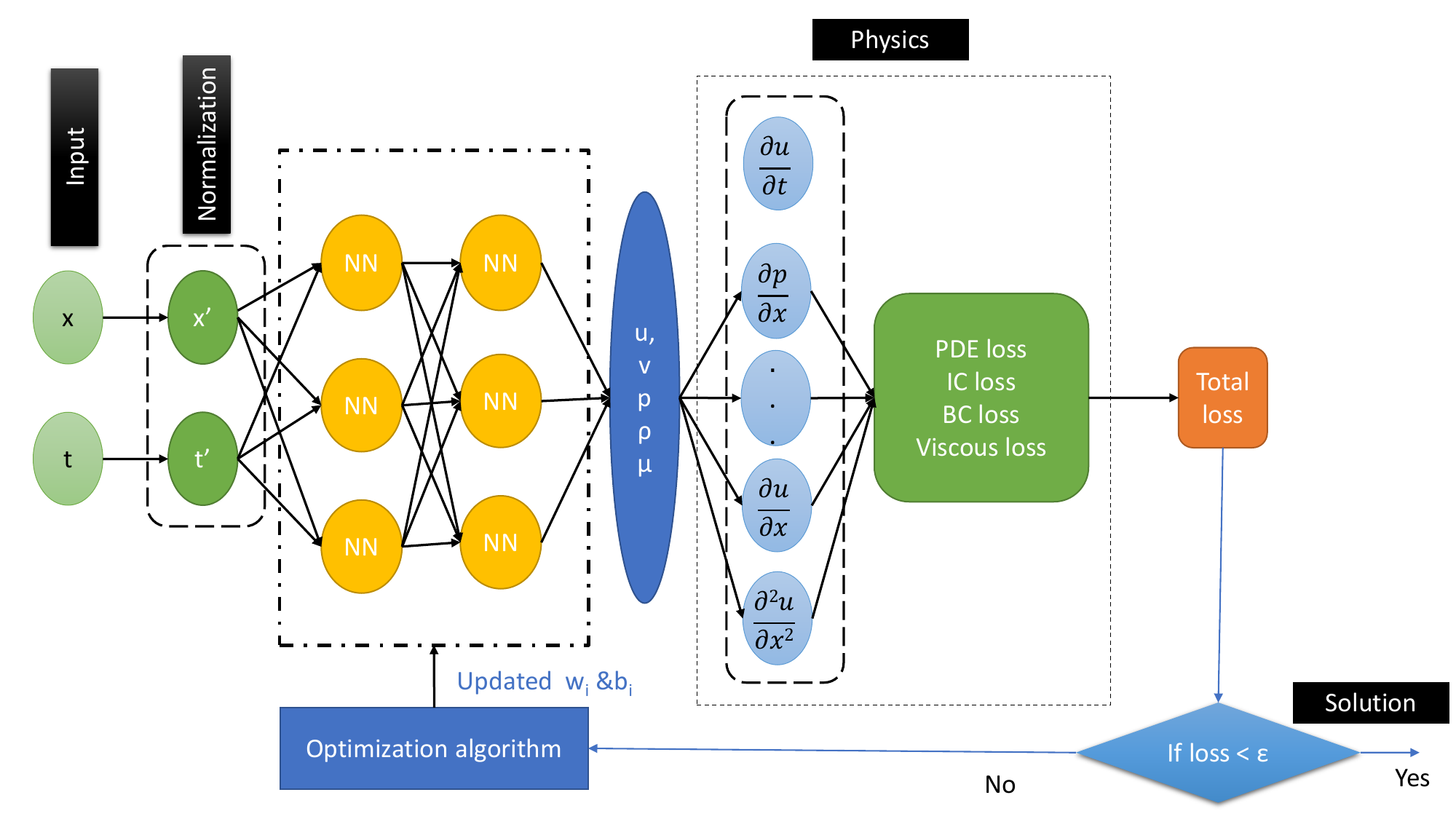}
	\caption{Steps used in Adaptive weight and viscosity PINNs architecture }
	\label{fig:PINN-AWV}
\end{figure}

\begin{itemize}
	\item Adaptive weights help to reduce the magnitude of PDE losses across the shocks and stabilizes the solver.
	\item Adaptive viscosity will try to find the optimal viscosity which stabilizes the solver.
\end{itemize}
Figure~\ref{fig:PINN-AWV} shows the steps used in the PINNs-AWV architecture. 

\section{Results}
In the previous sections we found numerical methods need are very sensitive to governing equation and discretization. In this section, we shall study whether PINNs also exhibit similar issue by testing it on few problems where numerical methods shows different results for conservative and non-conservative forms. To make a fair comparison we have included the non-conservative equation with artificial viscosity for problem having shocks.

\subsection{Shallow Water Equations with Variable Bottom Surface}

The one-dimensional shallow water equations with variable bathymetry that cannot be written in ``purely" conservative form. We already discussed in subsection~\ref{sec:swl}, where traditional numerical schemes need well-balanced property to accurately solve this problem. We have used the same discretization we used in subsection~\ref{sec:swl} for this problem. 
A reference solution is computed using a conservative finite-volume method implemented in PyClaw~\cite{pyclaw-sisc}. The equations are solved in conservative variables $(h, hu)$ using a Roe-type approximate Riemann solver combined with an operator-split treatment of the bathymetry source term~\cite{leveque-george-2008}. Reflective boundary conditions are imposed at both ends of the domain.

The finite-volume method preserves total mass to machine precision,
\begin{equation}
	M(t) = \int_0^L h(x,t)\,dx = \text{const},
\end{equation}
while exhibiting a small, physically consistent decay of total mechanical energy due to numerical dissipation and limiter effects. The resulting solution serves as a reference benchmark for assessing the qualitative behavior and conservation properties of the physics-informed neural network formulation.
The physics-informed neural network (PINNs) directly approximates the solution fields $h(x,t)$ and $u(x,t)$. The governing equations are enforced in conservative form by minimizing the residuals of the balance laws at randomly sampled spatio-temporal collocation points.
In addition to the standard PDE, initial-condition, and boundary-condition losses, a global mass constraint is introduced to penalize deviations from total mass conservation,
\begin{equation}
	\mathcal{L}_{\text{mass}} =
	\left(
	\int_0^L h(x,t)\,dx -
	\int_0^L h(x,0)\,dx
	\right)^2.
\end{equation}
This term weakly enforces global conservation without introducing explicit numerical fluxes or discretization stencils.
Since the solution remains smooth, errors are quantified using an $L_2$ norm. To remove contamination by the mean water level, we compare the free-surface perturbation,
\begin{equation}
	\eta'(x,t) = h(x,t) + b(x) - H_0.
\end{equation}
The relative $L_2$ error at final time $T$ is defined as
\begin{equation}
	\| \eta'_{\text{PINN}} - \eta'_{\text{ref}} \|_{L^2}
	=
	\left(
	\frac{1}{L}
	\int_0^L
	\left( \eta'_{\text{PINN}}(x,T) - \eta'_{\text{ref}}(x,T) \right)^2 dx
	\right)^{1/2}.
\end{equation}
For shock-dominated problems considered in subsequent sections, $L_1$ norms are employed instead to avoid excessive sensitivity to localized discontinuities.

\begin{figure}
	\centering
	\begin{subfigure}[b]{0.49\textwidth}
		\includegraphics[width=\textwidth]{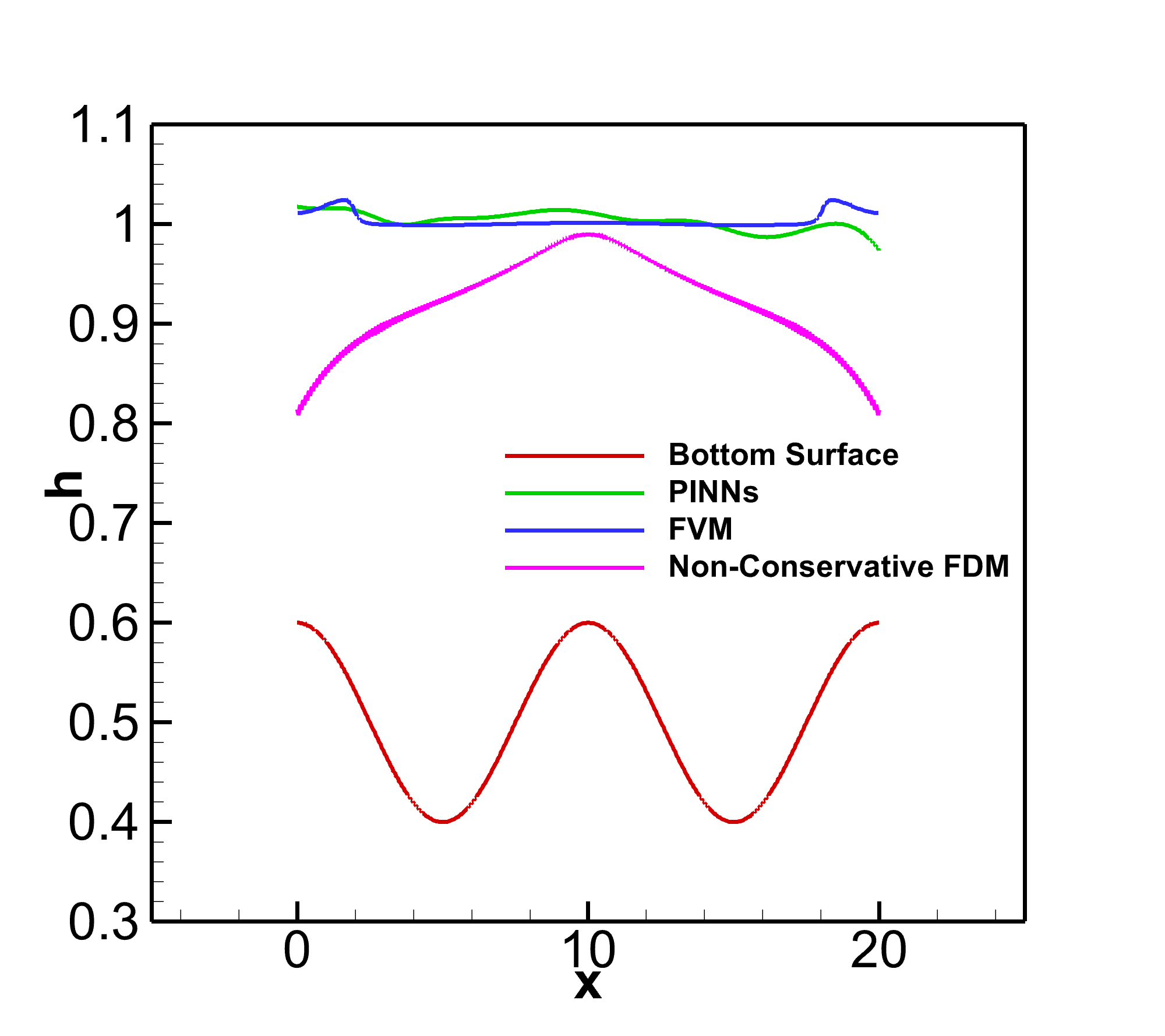}
		\caption{Free-surface elevation}
		\label{fig:h_SWL}
	\end{subfigure}
	\begin{subfigure}[b]{0.49\textwidth}
		\includegraphics[width=\textwidth]{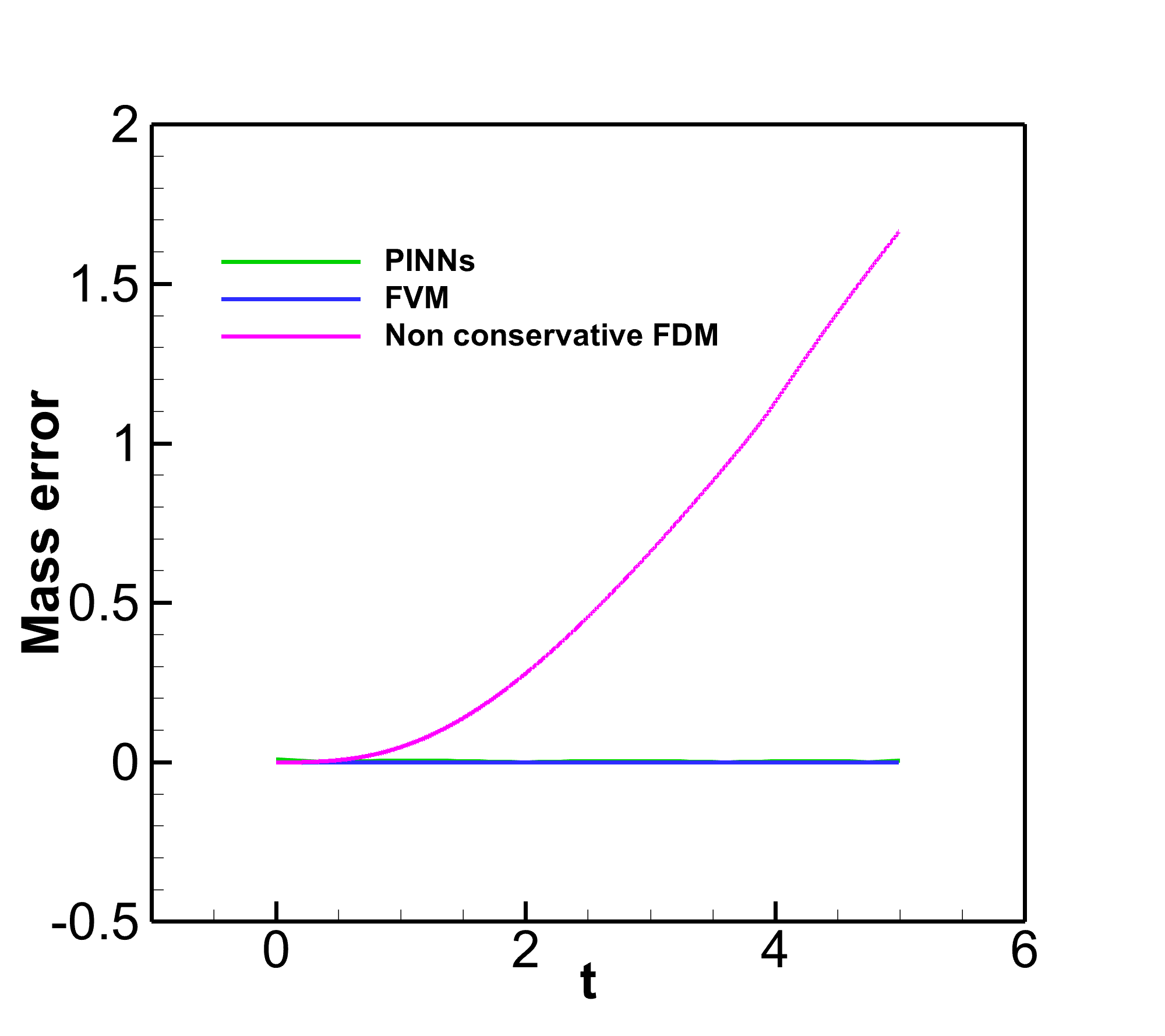}
		\caption{Mass error}
		\label{fig:mass_SWL}
	\end{subfigure}
	\caption{Solution of shallow water equation at t = 5.0 s with small initial disturbance}
	\label{fig:swl}
\end{figure}

Here, we have used the same discretization we used for subsection~\ref{sec:swl1}.
Figure~\ref{fig:swl} 
compares the free-surface elevation and mass conservation behavior obtained using a physics-informed neural network (PINNs), a conservative finite-volume method (FVM), and a non-conservative finite-difference method (FDM) for the shallow water equations with variable bathymetry.
Figure~\ref{fig:h_SWL}
shows the free-surface elevation at the final time 
t = 5.0 s. The conservative finite-volume solution closely follows the expected physical behavior, retaining small-amplitude oscillations associated with wave propagation and reflection over the variable bottom. The PINNs solution reproduces the large-scale structure of the free surface while exhibiting a smoother profile, reflecting the regularizing effect of optimization-based training. In contrast, the non-conservative finite-difference solution displays a significant distortion of the free surface, with an artificial amplification of the water level that is inconsistent with the underlying physics.

The PINNs solution reproduces the large-scale structure of the free surface while exhibiting a smoother profile, reflecting the regularizing effect of optimization-based training. \textbf{Notably, this result is obtained without the introduction of numerical fluxes, Riemann solvers, limiters, or specialized well-balanced discretizations. Instead, the PINNs relies solely on the governing balance laws and a global mass conservation constraint}, making the implementation conceptually simple compared to classical conservative solvers.

In contrast, the non-conservative finite-difference solution displays a significant distortion of the free surface, with an artificial amplification of the water level that is inconsistent with the underlying physics. This behavior is directly linked to the lack of a conservative formulation and illustrates the cumulative effect of local discretization errors on global invariants.
Figure~\ref{fig:mass_SWL} reports the evolution of the global mass error. The finite-volume method preserves mass to machine precision throughout the simulation, as expected from its conservative formulation. The PINNs solution also maintains bounded mass error without long-time drift, demonstrating that global conservation can be effectively enforced through integral constraints, despite the absence of explicit numerical fluxes. In sharp contrast, the non-conservative finite-difference scheme exhibits a monotonic growth in mass error, leading to a substantial accumulation of artificial mass over time. Table~\ref{tab:l2_error_comparison} shows the $L_2$ error of different scheme for this simulation. From this it is clear that PINNs solution is independent on the governing equation form for this problem.

These results highlight an important methodological distinction. While finite-volume methods achieve conservation through carefully designed numerical structures and problem-specific treatments, physics-informed neural networks can enforce the same global invariant through a comparatively simple optimization framework. At the same time, the smoother PINNs solution indicates that conservation alone is not sufficient to recover detailed hyperbolic wave dynamics, underscoring the complementary strengths and limitations of optimization-based and discretization-based approaches.

This experiment demonstrates that conservation of global invariants can be imposed in physics-informed neural networks without explicit flux discretization. However, conservation alone is not sufficient to guarantee accurate phase-resolved wave dynamics, underscoring a fundamental difference between numerical discretization and optimization-based solution strategies.
These findings motivate the subsequent investigation of shock-dominated problems, where the distinction between conservative and non-conservative formulations becomes even more pronounced.

\begin{table}[h]
	\centering
	\caption{Comparison of $L_2$ error in free surface elevation for different methods for SWE with no initial disturbance.}
	\label{tab:l2_error_comparison}
	\begin{tabular}{l c}
		\hline
		\textbf{Method} & \textbf{$L_2$ Error (Free Surface Elevation)} \\
		\hline

		\multicolumn{2}{c}{\emph{Numerical Methods}} \\
		 Non-Conservative Scheme & $9.2899 \times 10^{-2}$ \\
		Conservative Scheme & $1.0097 \times 10^{-1}$ \\
		Conservative Scheme (WB) & $8.7101 \times 10^{-15}$ \\
				\hline
		\multicolumn{2}{c}{\textit{Physics-Informed Neural Networks}} \\
		PINNs (Non-Conservative) & $4.7702 \times 10^{-4}$ \\
		PINNs (Conservative) & $6.9405 \times 10^{-3}$ \\
		\hline
	\end{tabular}
\end{table}

\subsection{Burgers equation on  discontinuous initial condition}  
\begin{figure}[H]
	\centering
	\begin{subfigure}[b]{0.49\textwidth}
		\includegraphics[width=\textwidth]{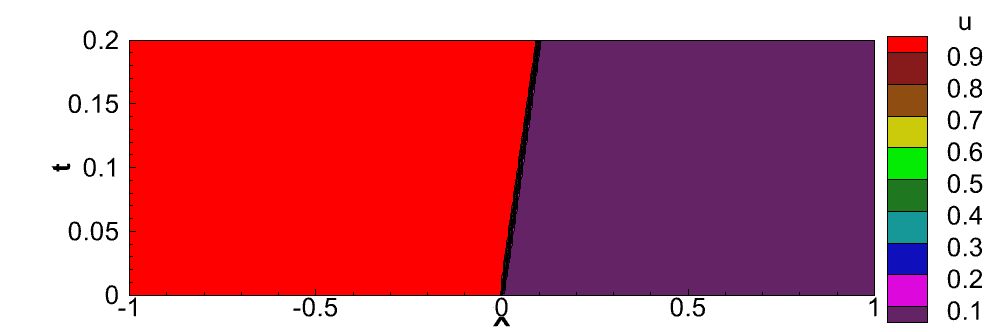}
		\caption{Conservative scheme}
		\label{fig:burg_con_num}
	\end{subfigure}
	\begin{subfigure}[b]{0.49\textwidth}
		\includegraphics[width=\textwidth]{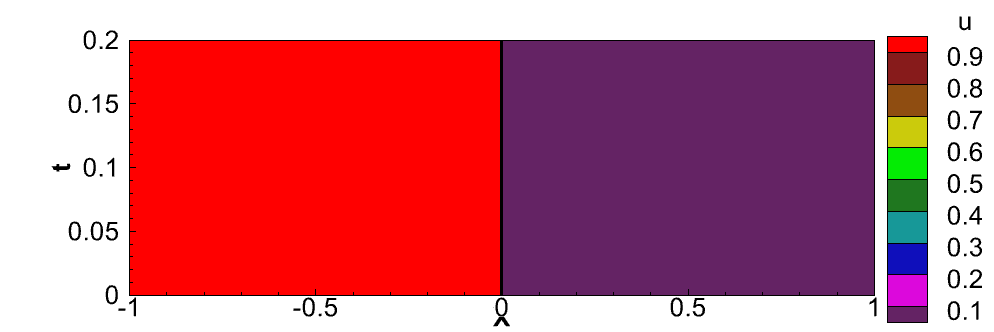}
		\caption{Non-conservative scheme}
		\label{fig:burg_non_con_num}
	\end{subfigure}
	
	\begin{subfigure}[b]{0.49\textwidth}
		\includegraphics[width=\textwidth]{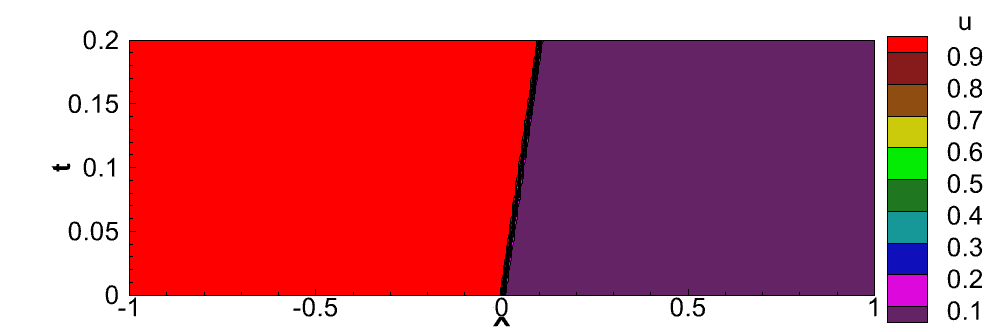}
		\caption{PINNs-conservative}
		\label{fig:burg_con_pinn}
	\end{subfigure}
	\begin{subfigure}[b]{0.49\textwidth}
		\includegraphics[width=\textwidth]{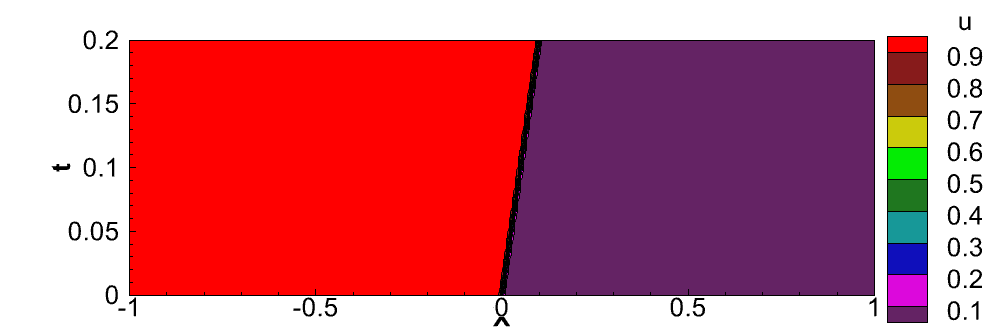}
		\caption{PINNs-non-conservative}
		\label{fig:burg_non_con_pinn}
	\end{subfigure}
	
	\caption{Solution of Burgers equation with discontinuous initial condition}
	\label{fig:bu2}
\end{figure}

\begin{figure}[H]
	\centering
\begin{subfigure}[b]{0.49\textwidth}
	\includegraphics[width=\textwidth]{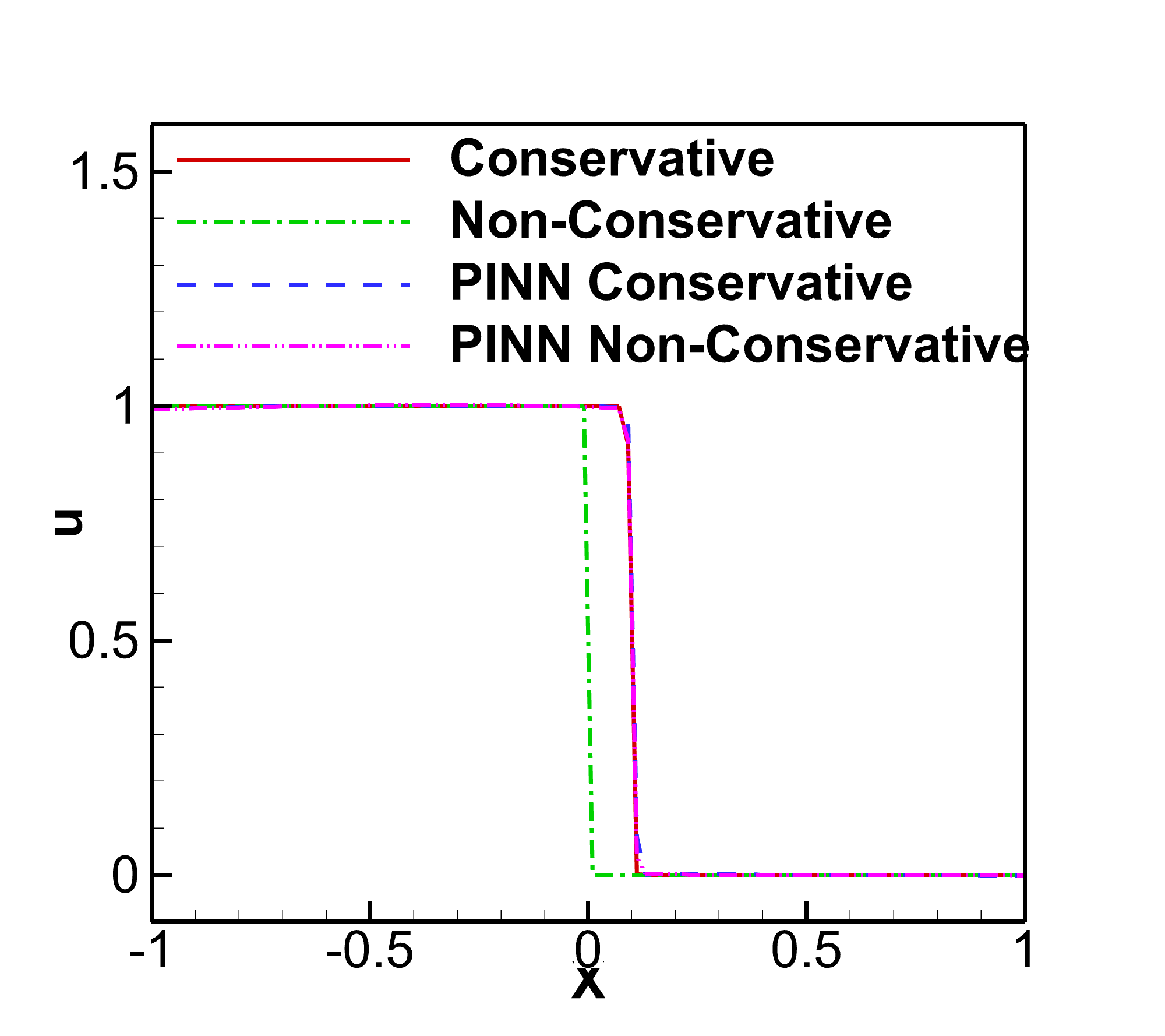}
	\caption{Numerical and PINNs}
	\label{fig:burg_d_p2_line}
\end{subfigure}
	\begin{subfigure}[b]{0.49\textwidth}
		\includegraphics[width=\textwidth]{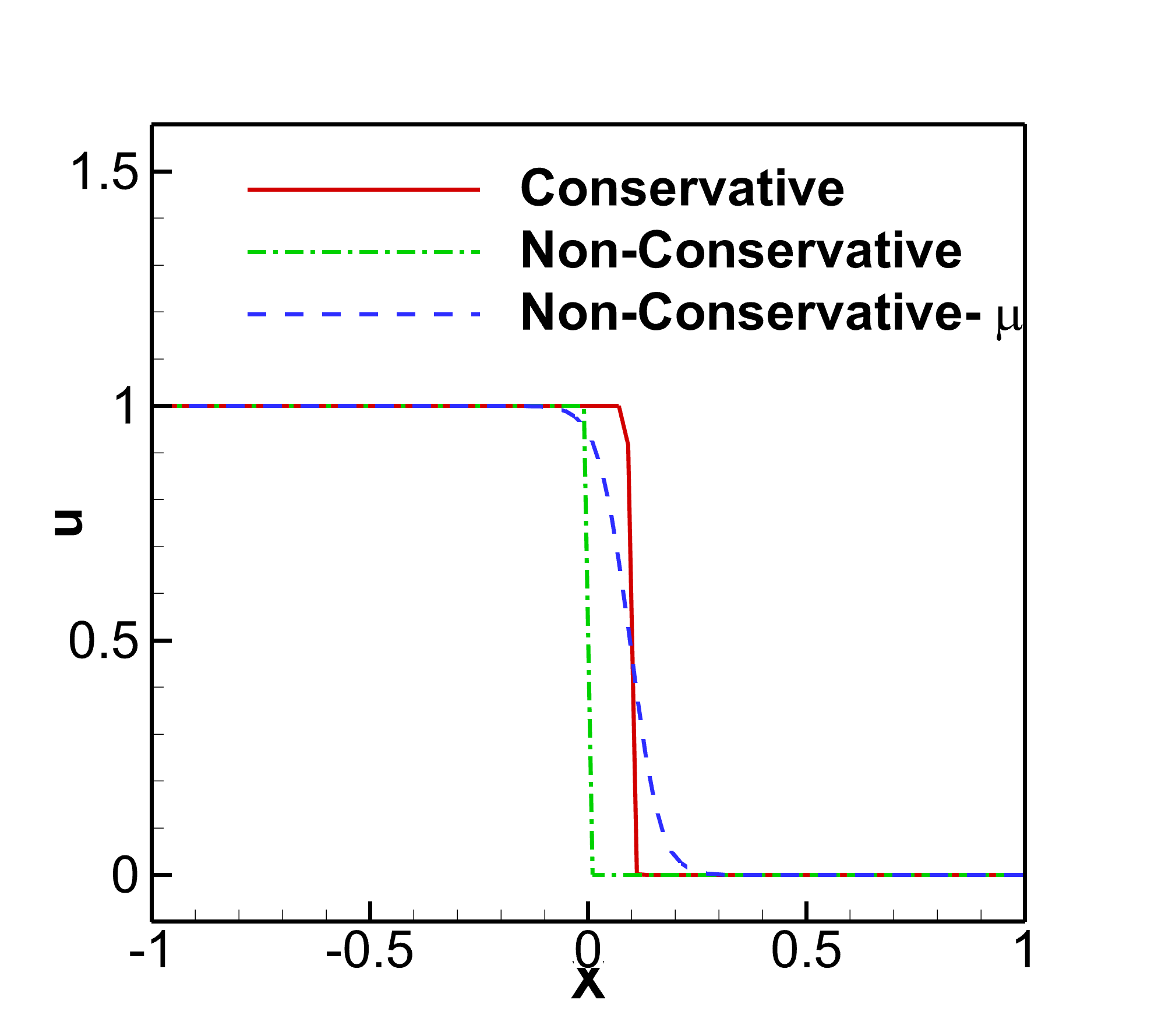}
		\caption{Numerical schemes }
		\label{fig:burg_d_p2_line1}
	\end{subfigure}
	\caption{Solution of Burgers equation at t = 0.2 s}
	\label{fig:bu3}
\end{figure}

In this test case, we have used a discontinuous initial condition, which is
\begin{equation}\label{key}
	u(x,0) = 
	\begin{cases}
		1.0, & \text{if } x \leq 0 \\
		0.0, & \text{if } x > 0
	\end{cases}
\end{equation}   
The same discretization procedure used in the subsection~\ref{sec:test1} is used here but we have used only 101 grid points for the simulation. The figure~\ref{fig:burg_con_num} shows the solution obtained using upwind scheme based on conservative discretization. Figure~\ref{fig:burg_non_con_num} shows the solution obtained using upwind scheme based on non-conservative discretization. From this, we can observe that the solution is \textit{not propagating} when we use non-conservative discretization on discontinuous initial condition. Figure~\ref{fig:burg_con_pinn} and \ref{fig:burg_non_con_pinn} shows the solution obtained using PINNs using the strong form of the conservative and non-conservative equations. From this, we can observe that PINNs produced the same result regardless of the form of the governing equation we used. Figure~\ref{fig:burg_d_p2_line} shows the solution of this test case at t=0.2. From this, we can observe that numerical methods based on a non-conservative scheme are unable to predict the shock location.   

We already studied that researchers used the artificial viscosity procedure to obtain the solution. So we added a scalar dissipation value of 0.001 and solved the problem. Figure~\ref{fig:burg_d_p2_line1} shows the solution of the Burgers equation solved with a scalar viscosity value of 0.001 using the non-conservative form of the Burgers equation. The solution at t=0.12 is shown in figure~\ref{fig:burg_d_p2_line1}. The non-conservative scheme is unable to predict shock speed when we don't use artificial viscosity, but is able to predict the shock speed when we add artificial viscosity. The magnitude of the viscosity and time step used is depends on the number of grid points and the initial condition we used. Though the non-conservative scheme with viscosity is able to predict the shock location, the scheme is very diffusive. When we reduce the magnitude of the viscosity, the scheme may become unstable. But in PINNs, it automatically find the optimal viscosity required to stabilize the solver. $L_1$ error and mass defect calculated for different scheme with respect to the exact solution is tabulated in table~\ref{tab:Burg1}.The results indicate that non-conservative formulations whether implemented via PINNs or traditional numerical discretizations—produce substantially larger errors than conservative schemes. While the non-conservative PINNs shows higher error relative to conservative approaches, it nevertheless captures the shock location effectively and achieves lower error than the viscously stabilized non-conservative finite-difference method.     

\begin{table}[h!]
	\centering
	\caption{Final-time diagnostics for Burgers' equation at $t = 0.2$.}
	\begin{tabular}{lcc}
		\hline
		Scheme & $L_1$ Error & Mass Defect \\
		\hline
		\multicolumn{3}{c}{\emph{Numerical Methods}} \\
		Conservative & $1.7226 \times 10^{-3}$ & $-1.5401 \times 10^{-14}$ \\
		Non-Conservative & $1.0101 \times 10^{-1}$ & $-1.0101 \times 10^{-1}$ \\
		Non-Conservative-$\mu$ & $4.6573 \times 10^{-2}$ & $-6.8782 \times 10^{-3}$ \\
		\hline
		\multicolumn{3}{c}{\emph{Physics-Informed Neural Networks}} \\
		Conservative PINNs & $4.3307 \times 10^{-3}$ & $-1.8086 \times 10^{-5}$ \\
		Non-Conservative PINNs & $4.8519 \times 10^{-3}$ & $-1.8893 \times 10^{-3}$ \\
		\hline
	\end{tabular}
\label{tab:Burg1}
\end{table}

%

\subsection{Sod Shock Tube Problem}

To assess the robustness of PINNs-AWV, we consider the Euler equations, a set of coupled non-linear hyperbolic equations. In this study, we examine the system in both conservative and non-conservative forms. The efficacy of these formulations has been demonstrated in previous literature for the Sod shock tube \cite{neelan2023efficient, neelan2021three} and supersonic wedge flow problems \cite{govind2022higher}.
The Euler equations in conservative form are given by:
\begin{equation}
	\frac{\partial}{\partial t}
	\begin{bmatrix}
		\rho \\
		\rho u \\
		E
	\end{bmatrix}
	+
	\frac{\partial}{\partial x}
	\begin{bmatrix}
		\rho u \\
		\rho u^2 + p \\
		u(E + p)
	\end{bmatrix}
	= 0
\end{equation}

where $\rho$, $u$, $p$, and $E$ denote density, velocity, pressure, and total energy per unit volume ($E = \frac{p}{\gamma - 1} + \frac{1}{2} \rho u^2$, $\gamma = 1.4$). The initial conditions for the Sod shock tube problem are:
\begin{equation}
	(\rho, u, p)(x, 0) =
	\begin{cases}
		(1.0,\ 0.0,\ 1.0), & \text{if } x < 0.5 \\
		(0.125,\ 0.0,\ 0.1), & \text{if } x > 0.5
	\end{cases}
\end{equation}

The non-conservative form is expressed as:
\begin{align*}
	\frac{\partial \rho}{\partial t} + u \frac{\partial \rho}{\partial x} + \rho \frac{\partial u}{\partial x} &= 0 \\
	\frac{\partial u}{\partial t} + u \frac{\partial u}{\partial x} + \frac{1}{\rho} \frac{\partial p}{\partial x} &= 0 \\
	\frac{\partial p}{\partial t} + u \frac{\partial p}{\partial x} + \gamma p \frac{\partial u}{\partial x} &= 0
\end{align*}

\begin{figure}
	\centering
	\begin{subfigure}[b]{0.45\textwidth}
		\includegraphics[width=\textwidth]{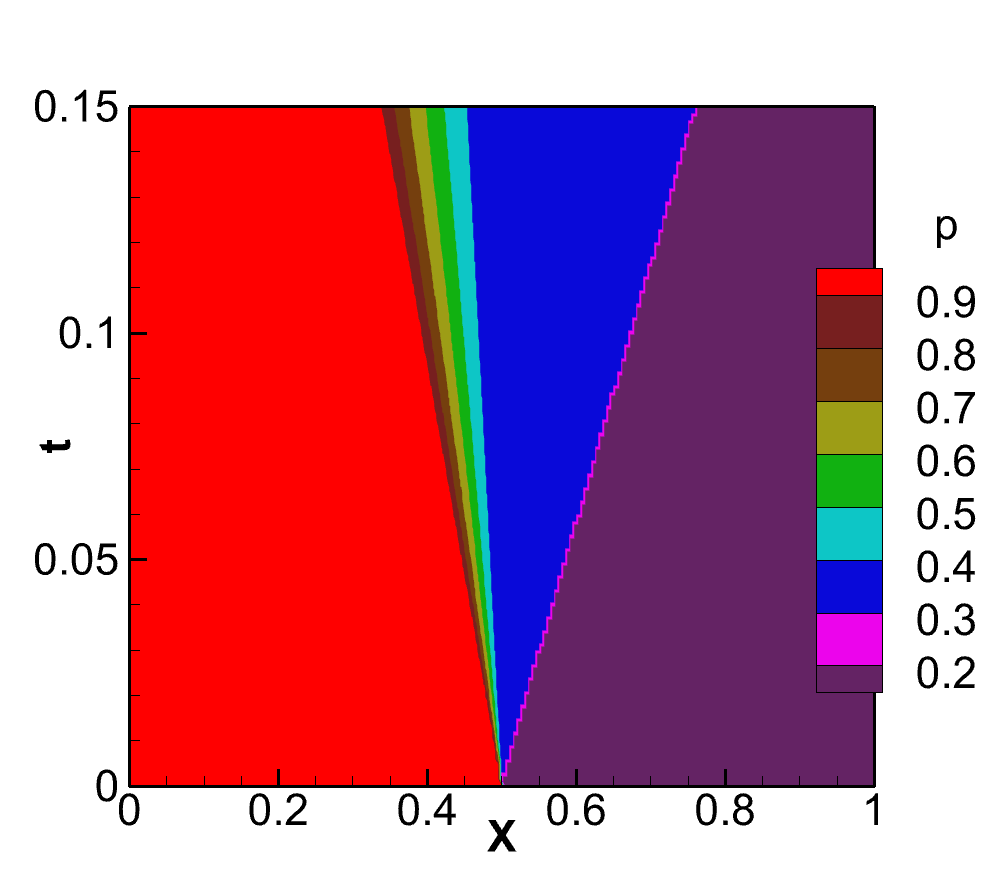}
		\caption{Exact solution}
		\label{fig:sod_exact}
	\end{subfigure}	
	\begin{subfigure}[b]{0.45\textwidth}
		\includegraphics[width=\textwidth]{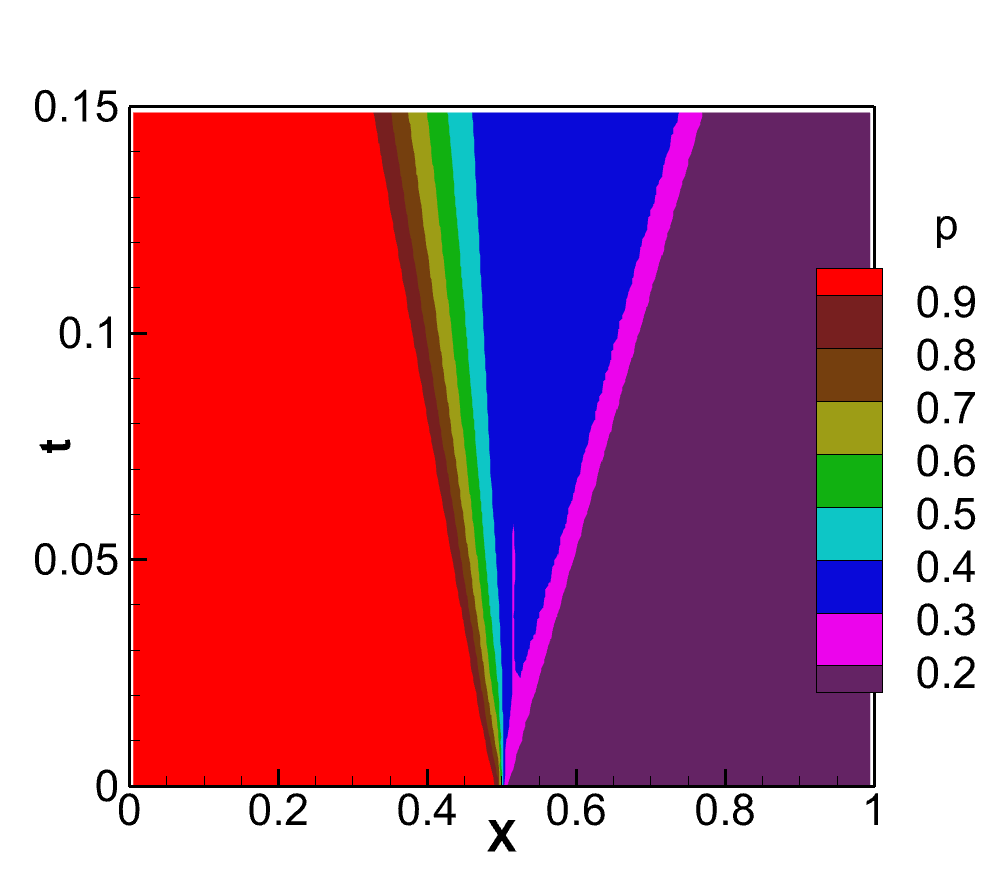}
		\caption{Conservative numerical scheme}
		\label{fig:Sod_Muscl}
	\end{subfigure}
	\begin{subfigure}[b]{0.45\textwidth}
		\includegraphics[width=\textwidth]{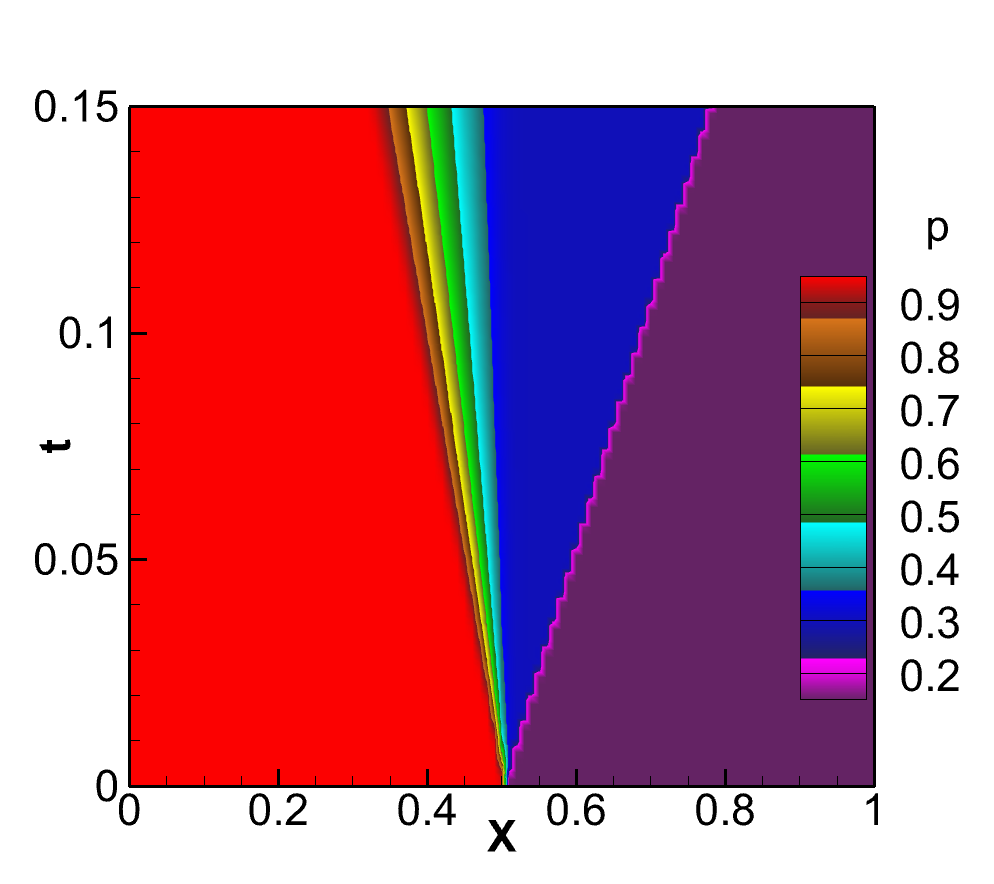}
		\caption{PINNs on Conservative equations}
		\label{fig:Sod_con_PINN}
	\end{subfigure}	
	\begin{subfigure}[b]{0.45\textwidth}
		\includegraphics[width=\textwidth]{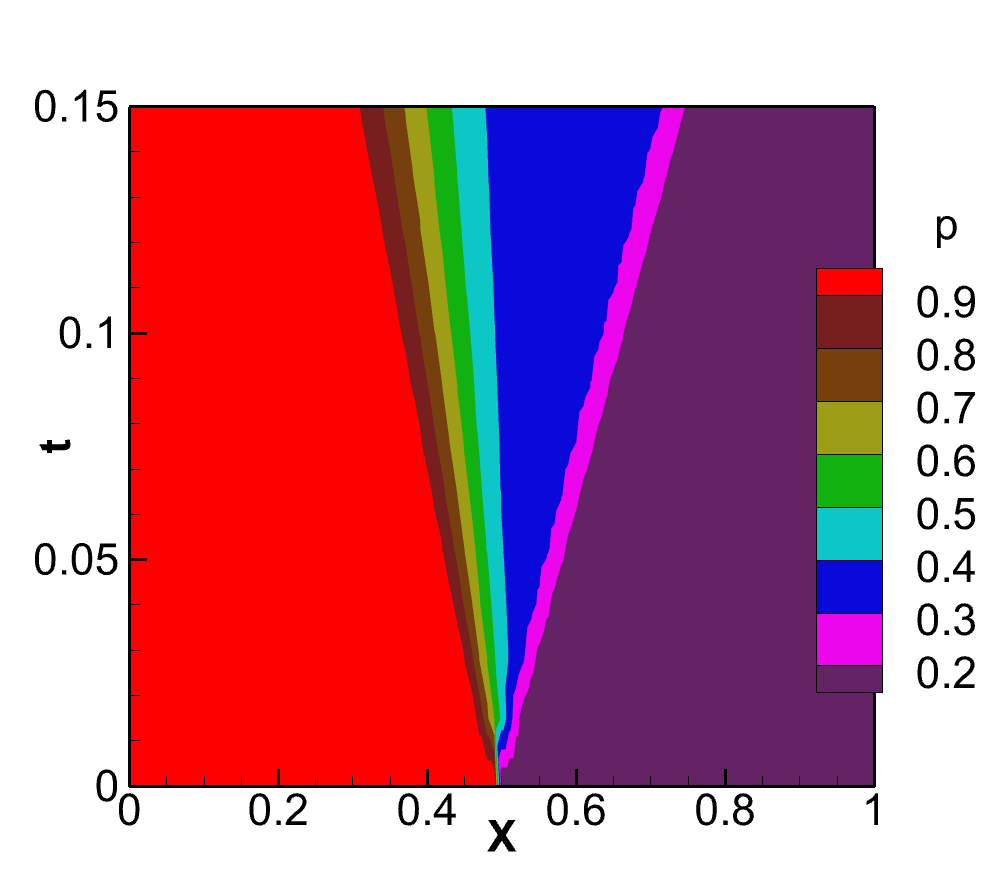}
		\caption{Non-conservative-$\mu$ numerical scheme}
		\label{fig:Sod_non_con_numerical}
	\end{subfigure}
	\begin{subfigure}[b]{0.45\textwidth}
		\includegraphics[width=\textwidth]{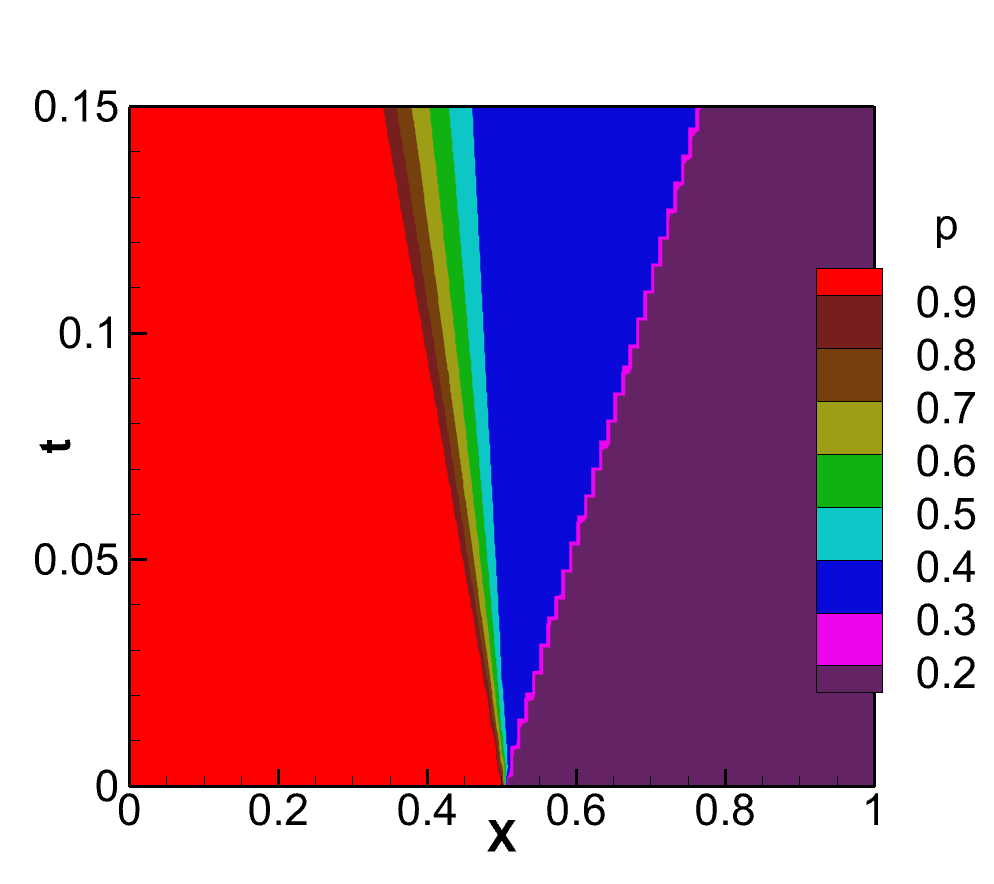}
		\caption{PINNs on non-conservative equations}
		\label{fig:Sod_non_con_PINN}
	\end{subfigure}
	\begin{subfigure}[b]{0.45\textwidth}
		\includegraphics[width=\textwidth]{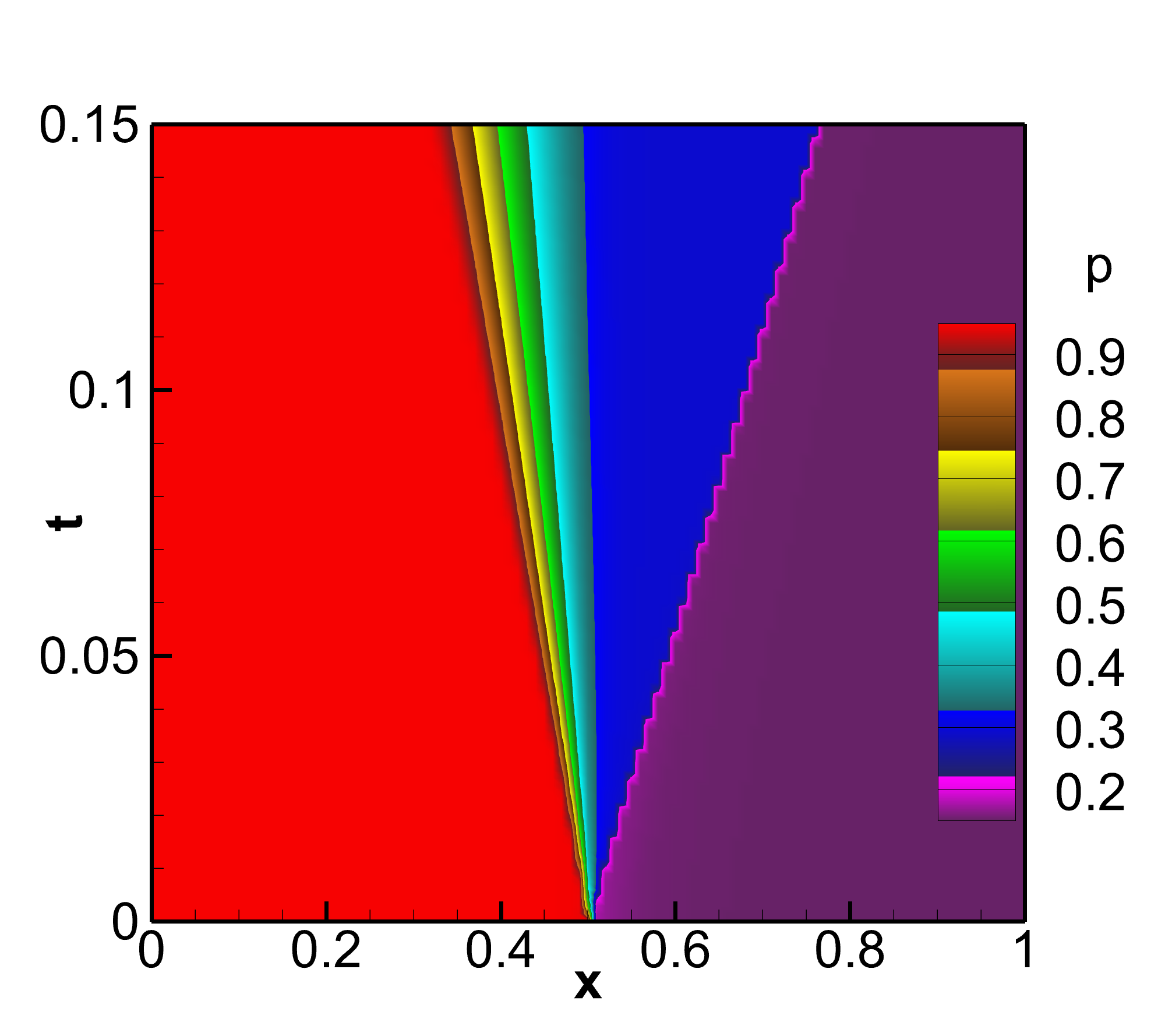}
		\caption{PINNs on non-conservative equations with path integral }
		\label{fig:Sod_non_con_PINNs_path}
	\end{subfigure}
	
	\caption{Pressure solution of Sod shock tube problem over time}
	\label{fig:sod}
\end{figure}

To handle this form numerically, we implement both a standard non-conservative- $\mu$ scheme and a path-conservative scheme based on Dal Maso-LeFloch-Murat (DLM) theory \cite{dal1995definition}.
The problem is solved using 100 uniform spatial and 100 temporal collocation points for PINNs-AWV \cite{neelan2024physics}. Numerical solutions are obtained on a 100-point grid using an HLLC solver, minmod limiter, and HRK time integration \cite{arun2018hyperbolic}.
The pressure contours are presented in Figure \ref{fig:sod}. While numerical methods resolve the shock (Figures \ref{fig:Sod_Muscl} and \ref{fig:Sod_non_con_numerical}), they exhibit significant diffusion. In contrast, the PINNs-AWV solutions (Figures \ref{fig:Sod_con_PINN}--\ref{fig:Sod_non_con_PINNs_path}) produce sharper profiles. Table \ref{tab:Sod_er} summarizes the $L_1$ errors and mass defect at $T = 0.12$ s. Notably, the numerical path-integral method shows improved mass conservation over the standard non-conservative scheme, yet PINNs maintain a lower error profile across the board at early flow times.

\subsubsection{Effect of PDE Formulation on Shock Speed in Adaptive-Viscosity PINNs}

Despite the success at early flow times, extended simulations reveal a fundamental limitation of non-conservative PINNs. Figure \ref{fig:press_Tp5} illustrates the pressure solution at $T = 0.5$ s. While PINNs applied to conservative PDEs capture the exact shock location, the non-conservative formulation fails to capture the correct shock speed. 
Interestingly, this failure is unique to system of equations; for the scalar Burgers equation ($u_t + uu_x = 0$), the non-conservative PINN accurately captures shock propagation even at $T = 4$ s (Figure \ref{fig:Burgers_longT}). This is because the analytical identity $u u_x = (0.5 u^2)_x$ holds for smooth neural approximations, making the formulations equivalent during training.
However, for the Euler system, regularizing the primitive-variable formulation ($V$) with adaptive viscosity $\nu$ introduces a non-conservative term when transformed back to conservative variables ($U$):
\begin{equation}
	U_t + F(U)_x = (\nu U_x)_x - \nu \frac{\partial^2 U}{\partial V^2}(V_x)^2
\end{equation}
This additional term modifies the effective viscous structure within the shock layer. As $\nu \to 0$, the solutions of the two systems may converge to different weak limits. This confirms that preserving conservative formulations is critical when applying PINNs to hyperbolic systems with discontinuities, as different regularizations of non-conservative systems converge to fundamentally different weak solutions \cite{dal1995definition}.

\subsubsection{Path-Integral Formulation for Non-Conservative PINNs}

A fundamental challenge in solving the Euler equations in the non-conservative (primitive) form $\mathbf{V} = [\rho, u, p]^T$ is that the product $\mathbf{A}(\mathbf{V})\mathbf{V}_x$ is not well-defined in the sense of distributions when shocks are present. To address this, we employ the Dal Maso-LeFloch-Murat (DLM) theory \cite{dal1995definition}, which defines non-conservative products across discontinuities by integrating along a family of paths $\psi(\tau; \mathbf{V}_L, \mathbf{V}_R)$ for $\tau \in [0, 1]$.

In this study, we utilize a linear path in the phase space:
\begin{equation}
	\psi(\tau) = \mathbf{V}_L + \tau(\mathbf{V}_R - \mathbf{V}_L)
\end{equation}
The total contribution of the non-conservative product across a localized gradient (the jump) is given by the path integral:
\begin{equation}
	\mathcal{P}(\mathbf{V}_L, \mathbf{V}_R) = \int_{0}^{1} \mathbf{A}(\psi(\tau)) \frac{\partial \psi(\tau)}{\partial \tau} d\tau
\end{equation}
where $\mathbf{A}(\mathbf{V})$ is the system Jacobian in primitive variables.

\subsubsection{Recovery of Correct Shock Speed via Path-Integral Formulation}

Figure~\ref{fig:press_Tp5} illustrates this discrepancy at $T = 0.5$ s, where the non-conservative formulation with standard artificial viscosity diverges significantly from the exact solution. However, by employing the Path-Integral (PI-PINN) approach, we successfully recover the correct shock location and propagation speed, matching the accuracy of the conservative formulation.
This success is rooted in the treatment of the non-conservative product $\mathbf{A}(\mathbf{V})\mathbf{V}_x$. In classical theory \cite{dal1995definition}, the weak solution of a non-conservative system depends on the viscous regularization path. A standard neural network, when regularized by adaptive viscosity $\nu$, tends to satisfy:
\begin{equation}
	\mathbf{V}_t + \mathbf{A}(\mathbf{V}) \mathbf{V}_x = (\nu \mathbf{V}_x)_x .
\end{equation}
As shown in the previous analysis, transforming this back to conservative variables $\mathbf{U}$ introduces a source term $\nu \frac{\partial^2 \mathbf{U}}{\partial \mathbf{V}^2}(\mathbf{V}_x)^2$ that does not vanish in the shock layer as $\nu \to 0$. This term shifts the Rankine-Hugoniot jump conditions, leading to the observed shock speed error.
The Path-Integral approach resolves this by explicitly enforcing the jump condition through the Dal Maso-LeFloch-Murat (DLM) framework. By integrating the Jacobian $\mathbf{A}(\mathbf{V})$ along a linear path in state space, the PI-PINN loss:
\begin{equation}
	Loss_{path} = \left\| \left( \int_{0}^{1} \mathbf{A}(\psi(\tau)) d\tau \right) \Delta \mathbf{V} - \Delta \mathbf{F} \right\|^2
\end{equation}
effectively constrains the neural network to satisfy the conservation properties of the system, even when expressed in primitive variables. Our results demonstrate that for the Sod shock tube problem at $T = 0.5$ s, the PI-PINN and the path-conservative numerical method both yield solutions that are consistent with the exact Riemann solution, effectively bridging the gap between non-conservative formulations and physical conservation laws. The numerical approximation of non-conservative hyperbolic systems has been rigorously studied using the path-conservative framework introduced by Parés \cite{munoz2007godunov}, which generalizes conservative finite volume schemes and Riemann solvers to systems containing non-conservative products. 
However, subsequent studies have shown that even when the theoretically correct path is used, path-conservative schemes may still produce incorrect shock solutions due to discretization errors and numerical viscosity effects \cite{abgrall2010comment}.

 \subsubsection{Implementation in PINNs-AWV}
\begin{figure}[h]
	\centering
	\begin{subfigure}[b]{0.49\textwidth}
		\includegraphics[width=\textwidth]{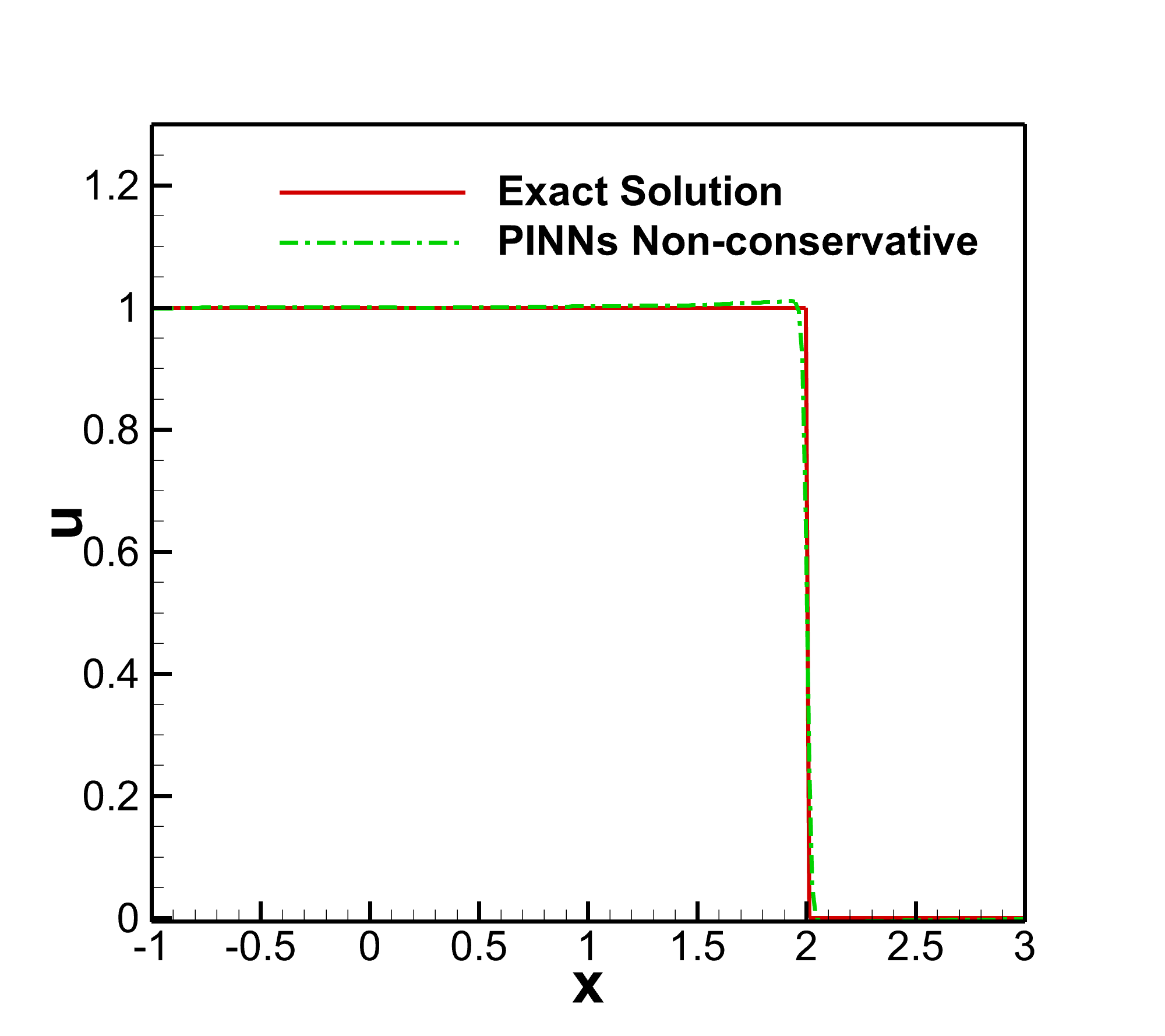}
		\caption{Burgers solution at T = 4 s}
		\label{fig:Burgers_longT}
	\end{subfigure}
	\begin{subfigure}[b]{0.49\textwidth}
		\includegraphics[width=\textwidth]{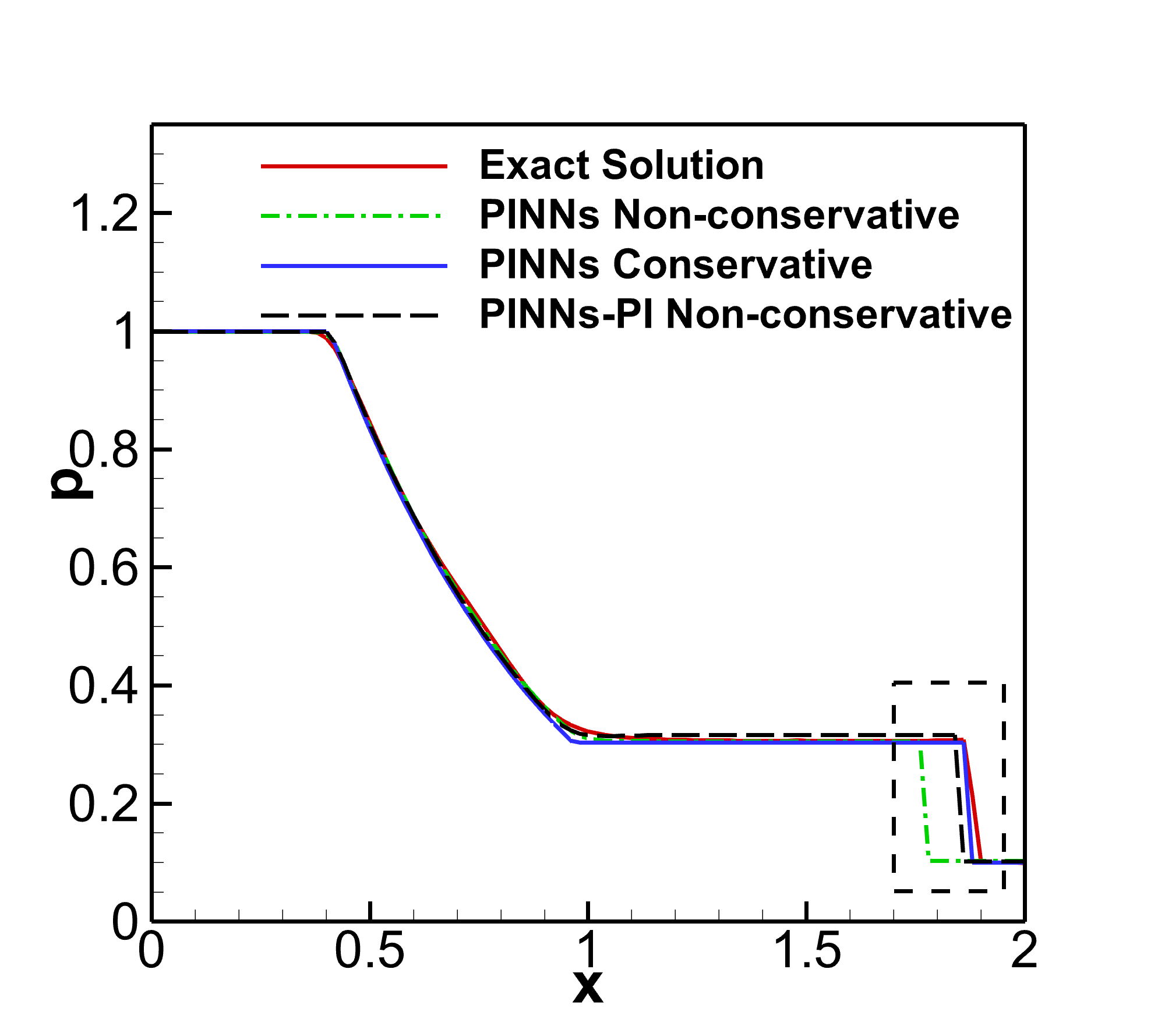}
		\caption{Pressure solution of Sod shock tube solution at T = 0.5 s}
		\label{fig:press_Tp5}
	\end{subfigure}
	
	\caption{Solution at long flow time }
	\label{fig:sodL}
\end{figure}

To implement this within the PINN framework, we modify the residual loss at each collocation point $(x, t)$. We define a localized stencil by perturbing the spatial coordinate by a small parameter $\epsilon$, yielding left and right states $\mathbf{V}_L = \mathbf{V}(x-\epsilon, t)$ and $\mathbf{V}_R = \mathbf{V}(x+\epsilon, t)$. The path integral is then approximated using 3-point Gauss-Legendre quadrature:
\begin{equation}
	\mathcal{I}_{path} \approx \sum_{i=1}^{3} \omega_i \mathbf{A}(\psi(\tau_i)) (\mathbf{V}_R - \mathbf{V}_L)
\end{equation}
where $\tau_i$ and $\omega_i$ are the quadrature nodes and weights, respectively. The Path-Integral loss ($Loss_{path}$) is defined as the squared difference between this integral and the conservative flux jump:
\begin{equation}
	Loss_{path} = \left\| \mathcal{I}_{path} - [\mathbf{F}(\mathbf{V}_R) - \mathbf{F}(\mathbf{V}_L)] \right\|^2
\end{equation}
This formulation ensures that the neural network approximation satisfies the generalized Rankine-Hugoniot conditions required for consistent shock capturing in non-conservative systems.
The pressure contours for the Sod shock tube problem are presented in Figure \ref{fig:sod}. We compare the results of PINNs-AWV (Conservative, Non-conservative, and Path-Integral) against traditional numerical schemes (Conservative FV, Non-conservative-$\mu$, and Path-conservative). 
At $T=0.12$ s, Table \ref{tab:Sod_er} shows that the Path-Integral PINN significantly improves the resolution of the shock compared to the standard non-conservative numerical method. Specifically, the numerical path-integral method achieves a mass defect of $2.48 \times 10^{-4}$, which is an order of magnitude improvement over the standard non-conservative-$\mu$ scheme ($-2.81 \times 10^{-3}$). While all PINN formulations show negligible differences in error at this early stage, the path-integral formulation provides a mathematically consistent framework for handling the primitive variables.
However, as flow time increases to $T=0.5$ s (Figure \ref{fig:press_Tp5}), the non-conservative formulations using artifical viscosity didn't produce correct shock speed but with path integral loss function, PINNs able to give correct shock speed.

\begin{table}[h]
	\centering
	\caption{Updated primitive-variable errors and mass defect for the Sod shock tube problem at T = 0.12 s.}
	\begin{tabular}{lcccc}
		\toprule
		Scheme & $L_1(\rho)$ & $L_1(u)$ & $L_1(p)$ & Mass Defect \\
		\midrule
		\multicolumn{5}{c}{\emph{Numerical Methods}} \\
		Conservative FV & $8.3393 \times 10^{-3}$ & $1.7016 \times 10^{-2}$ & $7.6215 \times 10^{-3}$ & $-6.0587 \times 10^{-13}$ \\
		Non-conservative-$\mu$ & $1.8629 \times 10^{-2}$ & $3.9520 \times 10^{-2}$ & $2.0391 \times 10^{-2}$ & $-2.8099 \times 10^{-3}$ \\
		Non-conservative-PI & $1.4340 \times 10^{-2}$ & $2.8636 \times 10^{-2}$ & $1.3445 \times 10^{-2}$ & $2.4829 \times 10^{-4}$ \\
		\midrule
		\multicolumn{5}{c}{\emph{Physics-Informed Neural Networks}} \\
		PINNs Conservative & $6.6526 \times 10^{-3}$ & $6.5373 \times 10^{-3}$ & $1.6889 \times 10^{-2}$ & $3.3238 \times 10^{-3}$ \\
		PINNs Non-conservative & $7.0354 \times 10^{-3}$ & $6.4271 \times 10^{-3}$ & $1.0955 \times 10^{-2}$ & $4.0261 \times 10^{-3}$ \\
		PINNs-PI Non-conservative & $1.7529 \times 10^{-2}$ & $3.7016 \times 10^{-2}$ & $1.9785 \times 10^{-2}$ & $-2.8099 \times 10^{-3}$ \\
		\bottomrule
	\end{tabular}
	\label{tab:Sod_er}
\end{table}

\begin{figure}[H]
	\centering
	\begin{subfigure}[b]{0.49\textwidth}
		\includegraphics[width=\textwidth]{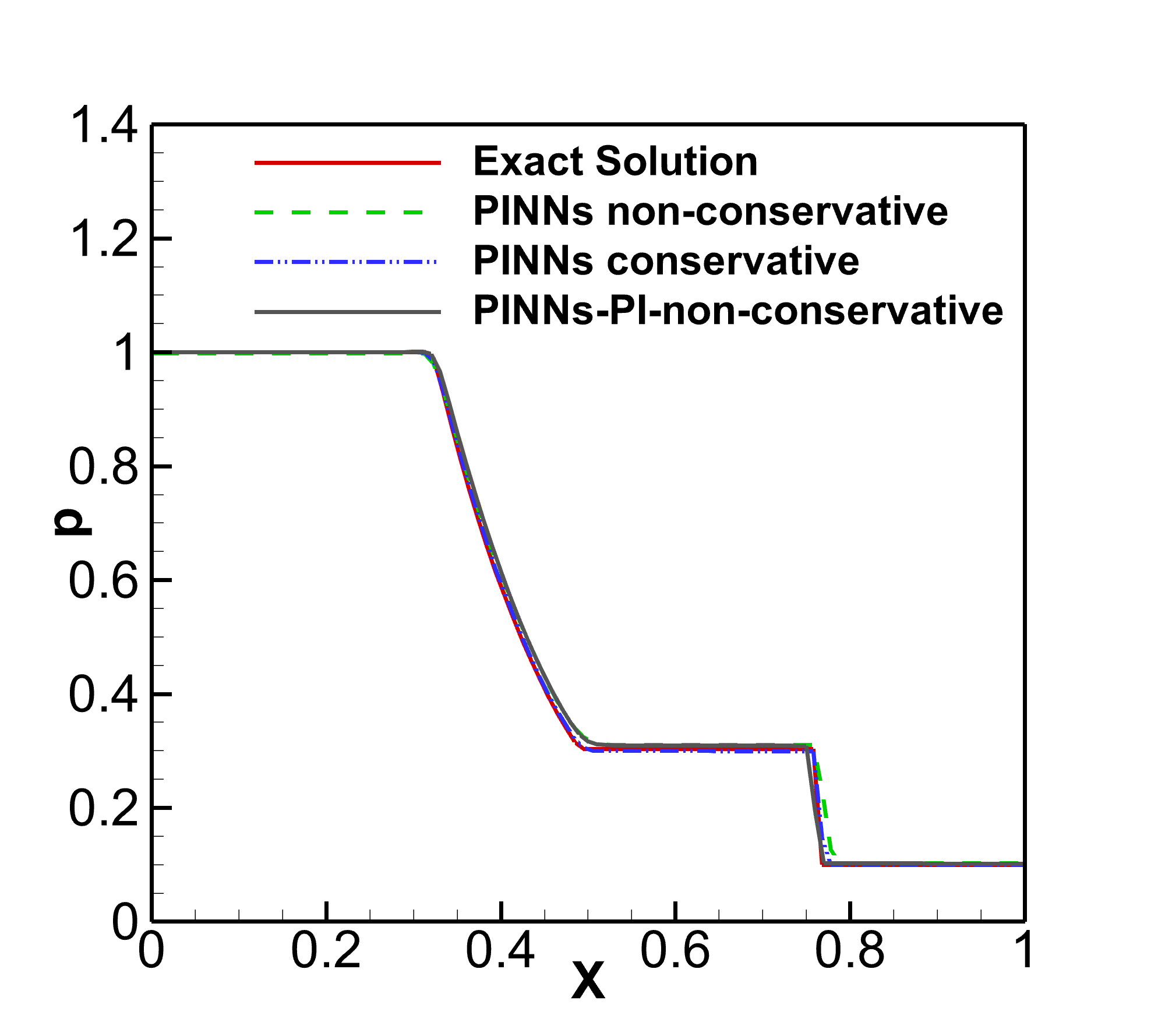}
		\caption{Using PINNs}
		\label{fig:Sod_line}
	\end{subfigure}
	\begin{subfigure}[b]{0.49\textwidth}
		\includegraphics[width=\textwidth]{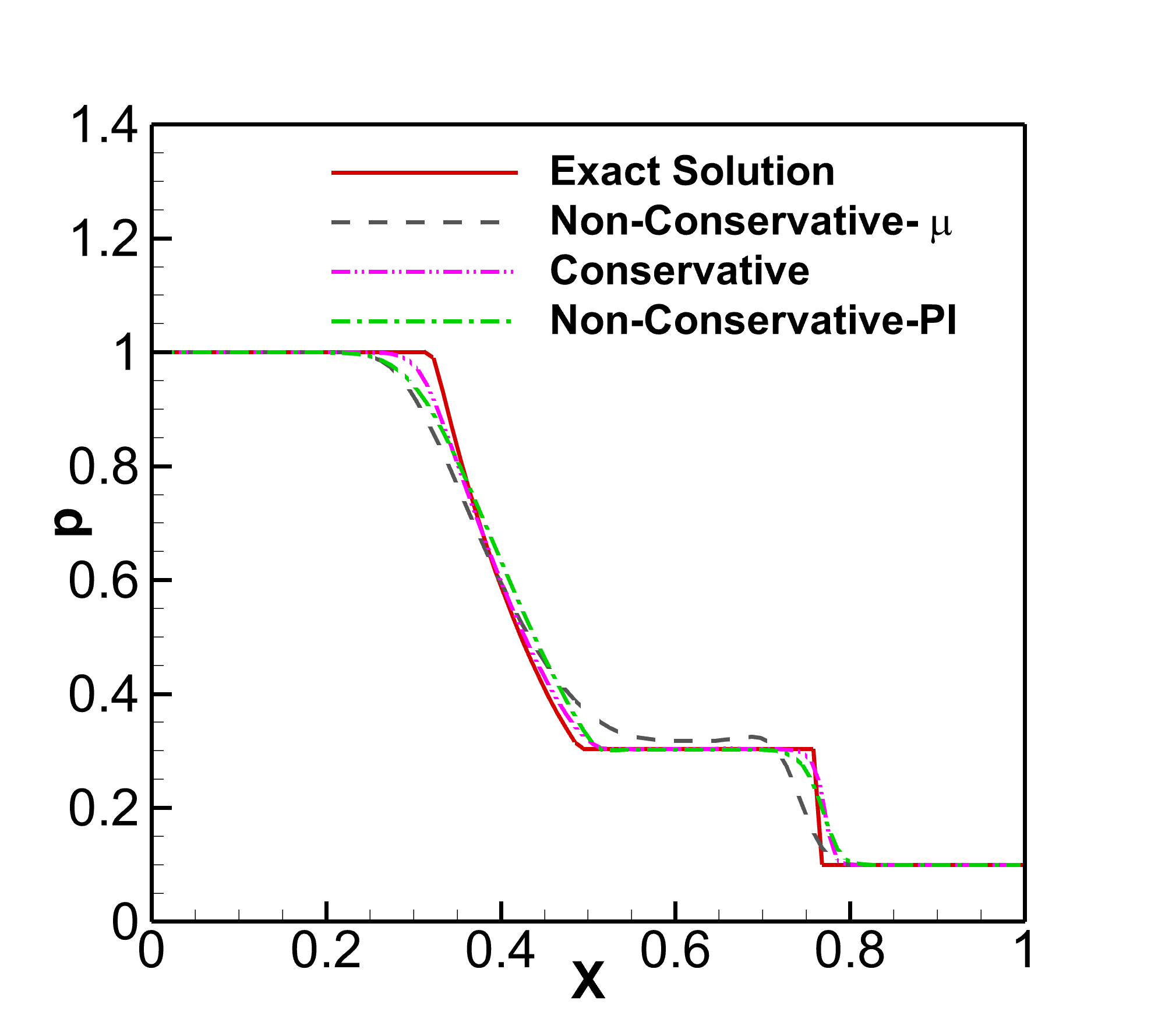}
		\caption{Using numerical methods}
		\label{fig:Sod_line}
	\end{subfigure}
	
	\caption{Pressure solution of Sod shock tube problem at T =0.15}
	\label{fig:sod}
\end{figure}

\begin{algorithm}[h]
	\caption{PINNs-AWV for Sod shock tube problem}
	\label{alg:pinn_sod}
	\begin{algorithmic}[1]
		\State Initialize PINNs Model
		\State Define Initial Condition Function $IC(x)$
		\State Generate spatio-temporal training data $x_{int\_train} = (t, x)$ for PDE residuals
		\State Generate initial condition data $x_{ic\_train} = (0, x)$ and corresponding $\rho_{ic\_train}, u_{ic\_train}, p_{ic\_train}$ using $IC(x)$	
		\Function{LossPDE}{$x_{int\_train}$}
		\State Predict $\rho, p, u, \nu_i = Neural Network(x_{int\_train})$
		\State Compute first derivatives: $\rho_t, \rho_x$, $u_t, u_x$, $p_t, p_x$ using 
		\State Compute second derivatives: $\rho_{xx}$, $u_{xx}$, $p_{xx}$
		\State Calculate adaptive weight term $d = 0.12 \cdot (|u_x| - u_x) + 1$
		\State Calculate viscosity $\nu = \nu_i^2$
		\State Formulate PDE residuals:
		\State $L_C = (\rho_t + u \rho_x + \rho u_x - \nu \rho_{xx}) / d$
		\State $L_M = (u_t + u u_x + (1 / \rho) p_x - \nu u_{xx}) / d$
		\State $L_E = (p_t + u p_x + \gamma p u_x - \nu p_{xx}) / d$
		\State \Return $\text{mean}(L_C^2) + \text{mean}(L_M^2) + \text{mean}(L_E^2) + \text{mean}(\nu^2)$
		\EndFunction
		
		\Function{LossIC}{$x_{ic\_train}, \rho_{ic\_train}, u_{ic\_train}, p_{ic\_train}$}
		\State Predict $\rho_{ic\_nn}, p_{ic\_nn}, u_{ic\_nn} = NeuralNetwork(x_{ic\_train})$
		\State \Return $\text{mean}((u_{ic\_nn} - u_{ic\_train})^2) + \text{mean}((\rho_{ic\_nn} - \rho_{ic\_train})^2) + \text{mean}((p_{ic\_nn} - p_{ic\_train})^2)$
		\EndFunction
		
		\State $L_{\text{pde}} = \text{LossPDE}(M, x_{int\_train})$
		\State $L_{\text{ic}} = \text{LossIC}(M, x_{ic\_train}, \rho_{ic\_train}, u_{ic\_train}, p_{ic\_train})$
		\State $L_{\text{total}} = L_{\text{pde}} + 1 \cdot L_{\text{ic}}$
		\State Compute gradients of $L_{\text{total}}$ with respect to model parameters
		\State \Return $L_{\text{total}}$	
		\State  \textbf{Phase 1: Adam Optimization}			
		\State  \textbf{Phase 2: L-BFGS Optimization}
		\State Trained model: $NeuralNetwork(t,x)$=$u$,$\rho$,$p$
		
	\end{algorithmic}
\end{algorithm}

\subsection{Supersonic flow over wedge}
The governing equations used are the steady-state two-dimensional Euler equations, which can be written in conservative differential form as:
\begin{equation}
	\frac{\partial F}{\partial x} + \frac{\partial G}{\partial y} = 0
\end{equation}
where the flux vectors are:
\begin{equation}
	F = \begin{pmatrix} \rho u \\ \rho u^2 + p \\ \rho uv \\ u(E+p) \end{pmatrix}, \quad G = \begin{pmatrix} \rho v \\ \rho uv \\ \rho v^2 + p \\ v(E+p) \end{pmatrix}.
\end{equation}

The total energy per unit volume is $E = \rho e + \frac{1}{2} \rho (u^2 + v^2)$. The equation of state for an ideal gas is $p = (\gamma - 1) \rho e$, which, expressed in terms of total energy $E$, is:
\begin{equation}
	p = (\gamma - 1) \left( E - \frac{1}{2} \rho (u^2 + v^2) \right).
\end{equation}

Alternatively, the non-conservative (primitive variable) form of the steady-state Euler equations can be expressed in terms of the primitive variable vector $W = [\rho, u, v, p]^T$ as:
\begin{equation}
	A \frac{\partial W}{\partial x} + B \frac{\partial W}{\partial y} = 0
\end{equation}
where the coefficient matrices $A$ and $B$ are given by:
\begin{equation}
	A = \begin{pmatrix} 
		u & \rho & 0 & 0 \\ 
		0 & u & 0 & \frac{1}{\rho} \\ 
		0 & 0 & u & 0 \\ 
		0 & \gamma p & 0 & u 
	\end{pmatrix}, \quad 
	B = \begin{pmatrix} 
		v & 0 & \rho & 0 \\ 
		0 & v & 0 & 0 \\ 
		0 & 0 & v & \frac{1}{\rho} \\ 
		0 & 0 & \gamma p & v 
	\end{pmatrix}.
\end{equation}

In this test case, supersonic flow with an inflow Mach number of 2 is simulated over a wedge with a $10^\circ$ angle. For the conservative formulation, we used Euler time integration with first-order reconstruction scheme with Lax-Fridrich scheme as Riemann solver. The goal of the present work is to study whether the non-conservative scheme can handle shocks not the accuracy.
For numerical methods we rotated the flow 10 degree so that we can use the advantage of Cartesian grid for this problem without special treatment for wall boundaries. We observed some mass defect in conservative scheme because we evaluate mass conservation by adding all inflow and outflow mass over the boundaries. Because we use transient simulation and didn't use the characteristic boundary condition we got non-zero mass defect for conservative scheme. 

\begin{figure}[htbp]
	\centering
	\begin{subfigure}[b]{0.49\textwidth}
		\includegraphics[width=\textwidth]{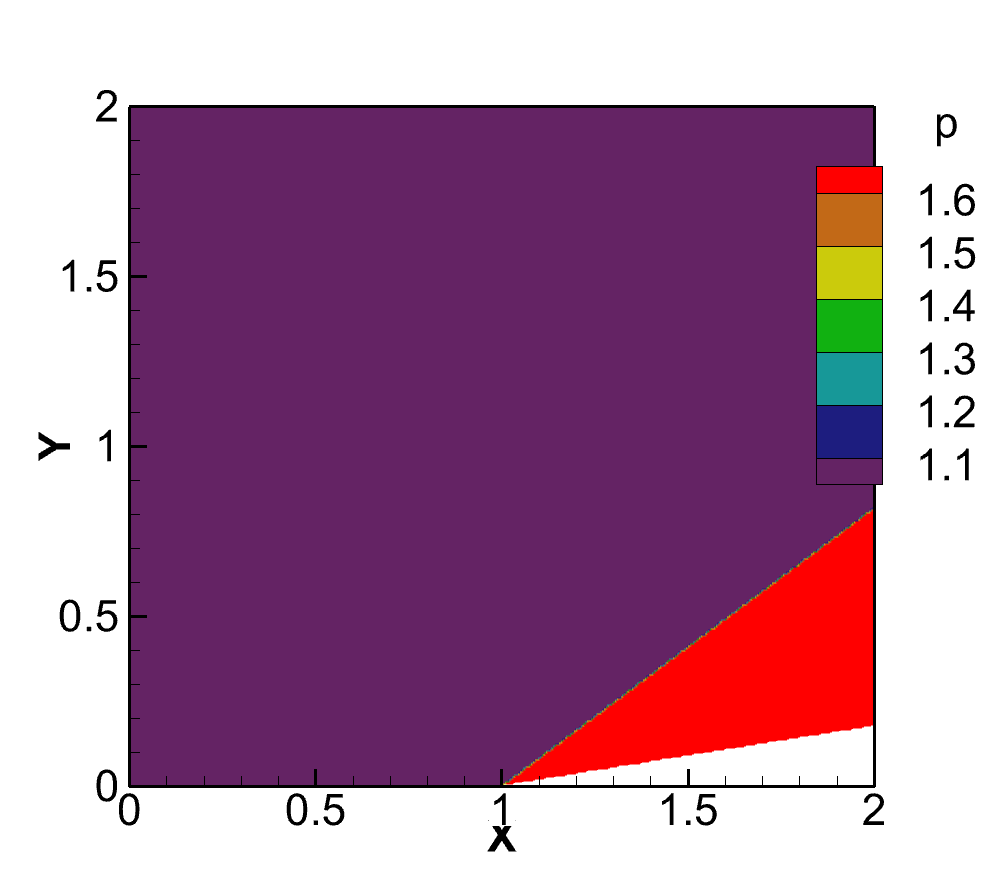}
		\caption{Exact solution}
		\label{fig:wedge_ex}
	\end{subfigure}
	\begin{subfigure}[b]{0.49\textwidth}
	\includegraphics[width=\textwidth]{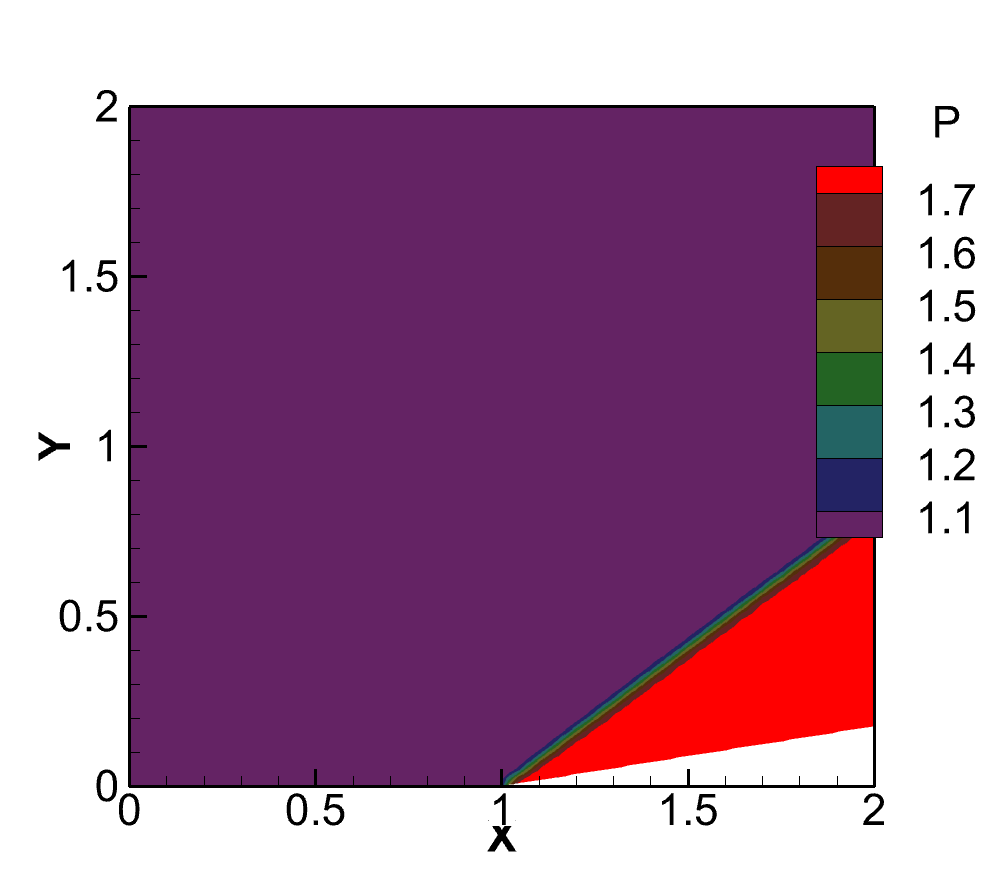}
	\caption{Conservative numerical method}
	\label{fig:Wedge_muscl_con}
\end{subfigure}
	\begin{subfigure}[b]{0.49\textwidth}
		\includegraphics[width=\textwidth]{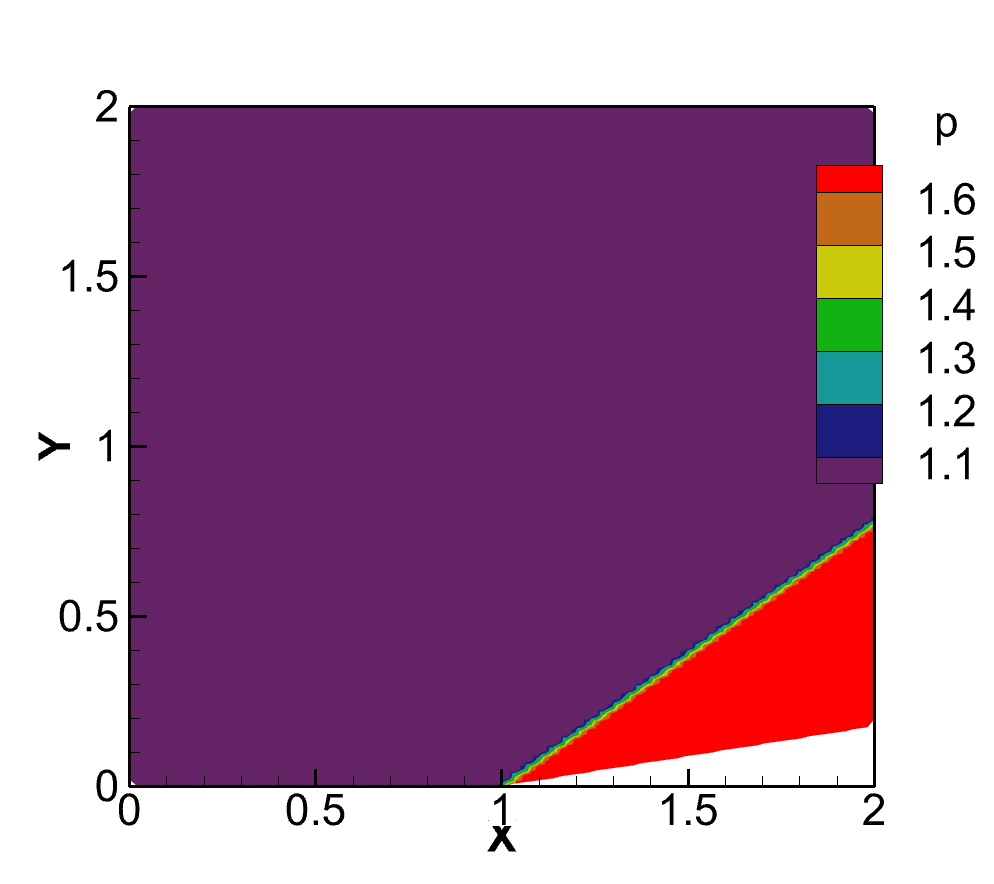}
		\caption{PINNs on Conservative equations}
		\label{fig:Wedge_pinn_con}
	\end{subfigure}	
	\begin{subfigure}[b]{0.49\textwidth}
		\includegraphics[width=\textwidth]{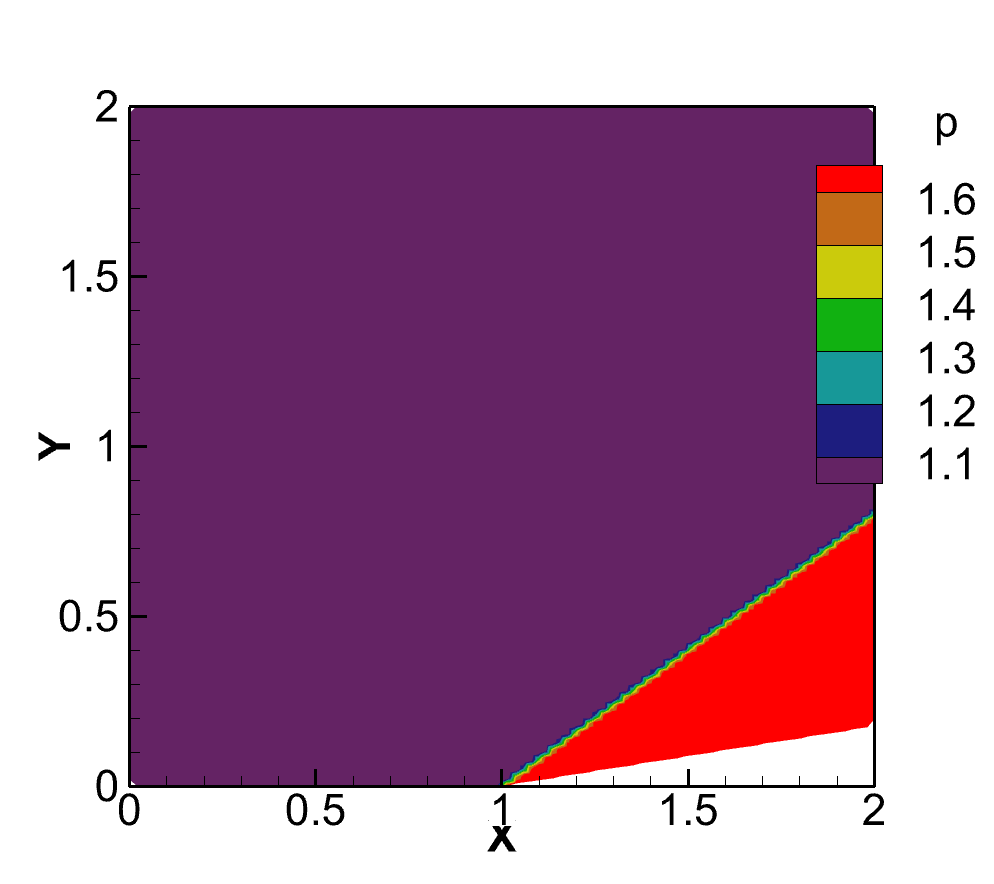}
		\caption{PINNs on non-conservative equations}
		\label{fig:Wedge_pinn_non_con}
	\end{subfigure}	
	\begin{subfigure}[b]{0.49\textwidth}
	\includegraphics[width=\textwidth]{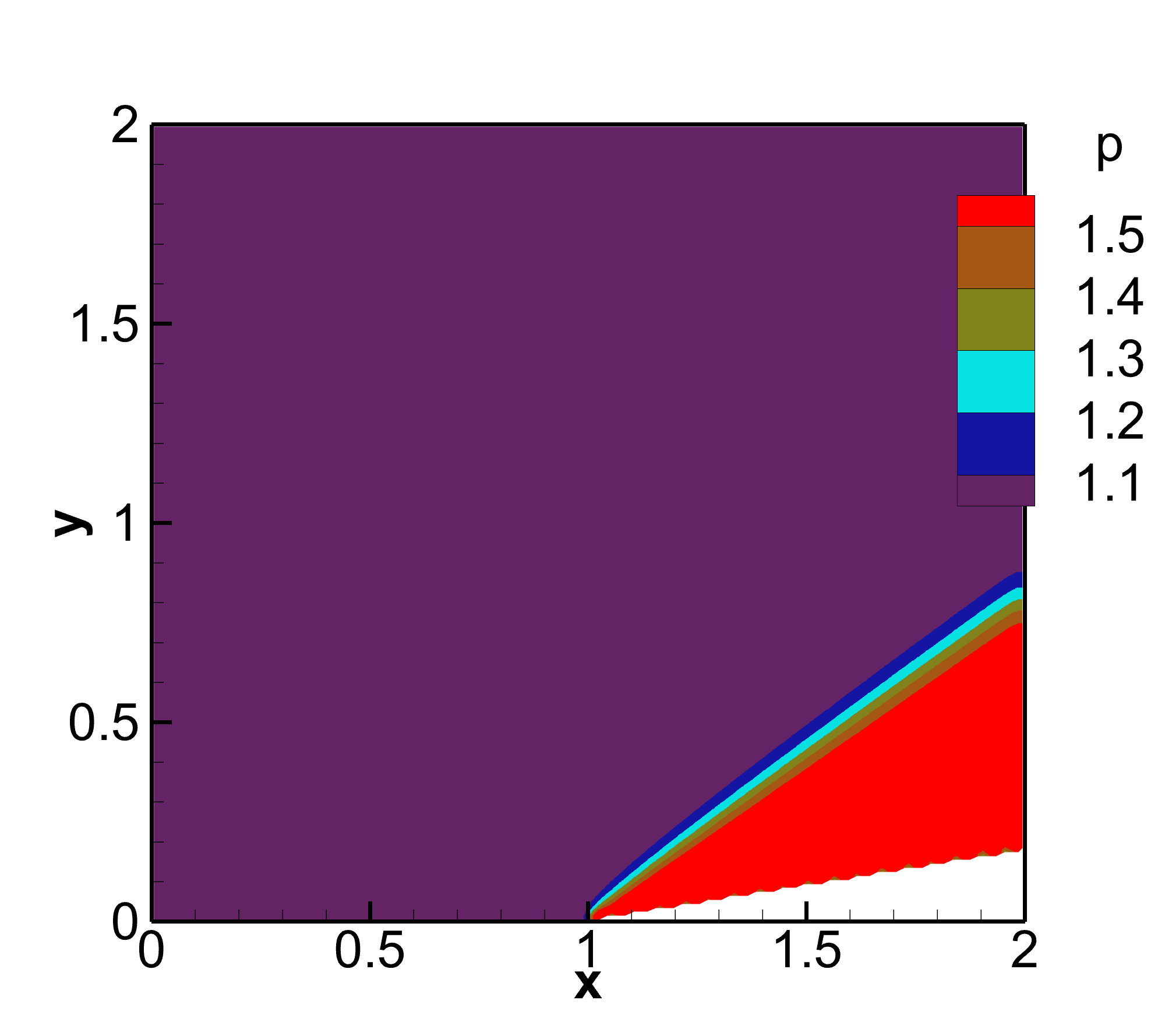}
	\caption{Non-Conservative-$\mu$ numerical method}
	\label{fig:wedge_non_con}
\end{subfigure}

	\caption{Pressure solution of supersonic flow over wedge}
	\label{fig:wedge}
\end{figure}

We employed 10,000 collocation points in the PINNs simulations, utilizing an adaptive-weight and viscosity-based PINNs architecture tailored for this problem. The exact solution is presented in Figure~\ref{fig:wedge_ex}. A slip boundary condition was applied along the wedge walls, with specified inflow parameters imposed at the inlet. 

Figures~\ref{fig:Wedge_pinn_con} and \ref{fig:Wedge_pinn_non_con} display the pressure contours obtained from PINNs simulations using the conservative and non-conservative forms of the governing equations, respectively. For comparison, Figure~\ref{fig:Wedge_muscl_con} shows the pressure contours computed with a conventional conservative numerical method employing the MUSCL scheme. More details about the numerical discretization can be found in~\cite{govind2022higher}.

\begin{figure}[htbp]
	\centering
	\includegraphics[width=0.7\linewidth]{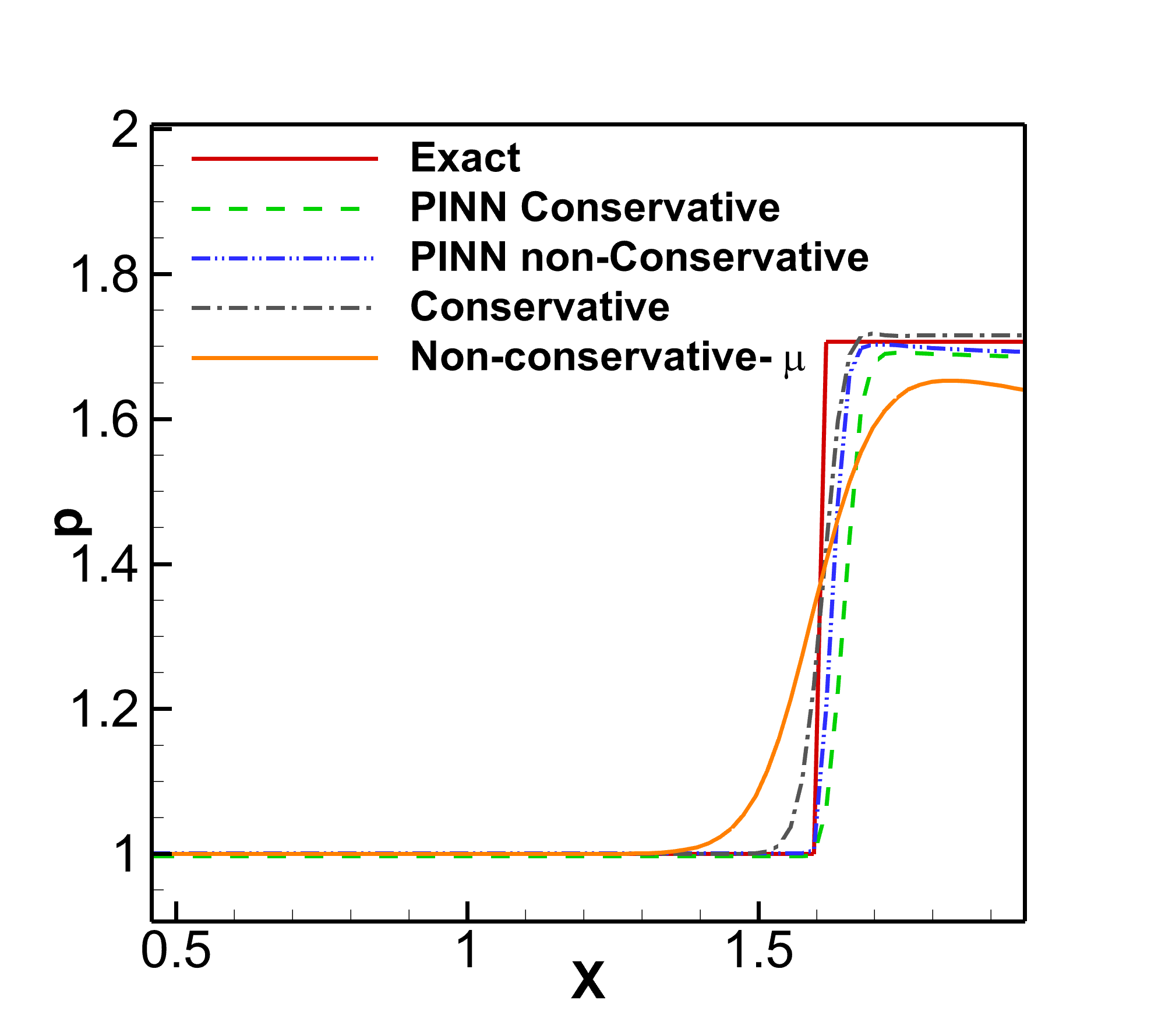}
	\caption{Pressure at y=0.5 in supersonic flow over wedge}
	\label{fig:wedge_line}
\end{figure}

In our PINNs simulations, shocks are resolved with approximately three grid points, consistent with the findings of Neelan et al.~\cite{neelan2024physics}. Notably, all methods successfully capture the shock structures. Furthermore, as demonstrated by the inclusion of the non-conservative Euler equations for PINNs in this comparison, the PINNs framework accurately captures the correct shock location for this problem without the positional discrepancies typically encountered in standard non-conservative schemes. 
Figure~\ref{fig:wedge_line} compares the line plots of pressure solutions obtained from PINNs-AWV simulations using both the conservative and non-conservative formulations of the Euler equations at $y = 0.5$ location. This demonstrates that the PINNs framework yields a consistent shock-resolving capability irrespective of the chosen form of the governing equations for steady case. Table~\ref{tab:wedge} quantitatively compares the $L_1$ error in pressure and the mass defect across the different methods. However, this localized accuracy comes at the cost of strict mass conservation; the non-conservative PINNs exhibits a mass defect of $3.39 \times 10^{-3}$, which is nearly an order of magnitude higher than that of the conservative PINNs ($3.62 \times 10^{-4}$). As expected, the FVM ensures perfect mass conservation.

\begin{table}[htbp]
	\centering
	\caption{Error and Mass Conservation for supersonic flow over wedge}
	\begin{tabular}{lcc}
		\hline
		Method & $L_1$ Error in $p$ & Mass Defect \\
		\hline
		PINNs (Conservative)      & 0.0202 & 3.62 $\times 10^{-4}$ \\
		PINNs (Non-conservative)  & 0.0849 & 3.39 $\times 10^{-3}$ \\
		Conservative FVM                      & 0.01150 &  5.84$\times 10^{-12}$\\
		Non-conservative-$\mu$ &0.11929& 6.29$\times 10^{-2}$\\
		\hline
	\end{tabular}
	\label{tab:wedge}
\end{table}

\section{Conclusion}

This study provides a comprehensive evaluation of conservative and non-conservative governing equations, tracing their performance from classical numerical solvers to physics-informed neural networks (PINNs) frameworks. By investigating the transition from scalar conservation laws to complex systems, we have identified the mechanisms that govern shock-capturing accuracy in neural-based solvers.
Our analysis of the Euler equations reveals that standard non-conservative partial differential equations (PDEs) consistently produce incorrect shock speeds when solved using both PINNs and traditional numerical methods—even when augmented with adaptive or artificial viscosity. These inaccuracies in shock propagation are not merely a result of discretization errors or insufficient resolution; rather, they stem from a fundamental structural deficiency: the inherent inability of standard non-conservative formulations to satisfy the Rankine-Hugoniot conditions across discontinuities. In steady-state problems, the solution is independent of shock propagation speed. As a result, even non-conservative formulations can predict correct shock locations. However, such formulations may introduce non-physical mass generation or loss due to the lack of discrete conservation. In the context of PINNs, this issue can be slightly mitigated by incorporating additional mass-conservation constraints into the loss function.

However, this study demonstrates that this limitation can be successfully overcome through the integration of a \textbf{Path-Integral (Path-Conservative) formulation} based on the Dal Maso-LeFloch-Murat (DLM) theory. By incorporating a path-consistent loss function that integrates the system Jacobian along a defined state-space path, we found that:
\begin{itemize}
	\item \textbf{Recovery of Physical Consistency:} The Path-Integral approach allows both numerical methods and PINNs to recover the mathematically correct shock speeds, effectively bridging the gap between primitive-variable formulations and physical conservation laws.
	\item \textbf{Resolution of Regularization Discrepancies:} The shock-speed drift observed in standard primitive-variable formulations is rooted in non-vanishing source terms introduced by viscous regularization. The path-integral constraint effectively nullifies these errors by ensuring the localized gradient satisfies the required jump manifold.
	\item \textbf{Neural Network Versatility:} PINNs-AWV demonstrates a unique capability to enforce these path-integral constraints globally. While conservative formulations remain the most straightforward for transient systems, the path-integral framework provides a robust alternative for applications where primitive variables are preferred or conservative forms are analytically complex.
\end{itemize}

Overall, this study establishes that conservativeness is not merely a property of the governing equations, but a hierarchical requirement that must be enforced at the mathematical, numerical, and physical levels. The proposed framework offers a principled pathway for designing physically consistent machine learning solvers for high-speed compressible flows.

\section*{Data and Code Availability}
The source code for this project is publicly available on GitHub at \url{https://github.com/AGN000/Conservative-And-Non-conservative-PINNs-paper}.

\bibliographystyle{unsrt}
\bibliography{sn-bibliography}

@book{toro2009riemann,
  title     = {Riemann Solvers and Numerical Methods for Fluid Dynamics: A Practical Introduction},
  author    = {Toro, Eleuterio F.},
  year      = {2009},
  edition   = {3rd},
  publisher = {Springer},
  address   = {Berlin, Heidelberg},
  isbn      = {978-3-540-25202-3},
  doi       = {10.1007/b79761}
}

@article{shapiro2006non,
  title={Non-conservative and conservative formulations of characteristics-based numerical reconstructions for incompressible flows},
  author={Shapiro, Evgeniy and Drikakis, Dimitris},
  journal={International journal for numerical methods in engineering},
  volume={66},
  number={9},
  pages={1466--1482},
  year={2006},
  publisher={Wiley Online Library}
}

@article{turkel1999preconditioning,
  title={Preconditioning techniques in computational fluid dynamics},
  author={Turkel, Eli},
  journal={Annual Review of Fluid Mechanics},
  volume={31},
  number={1},
  pages={385--416},
  year={1999},
  publisher={Annual Reviews}
}

@article{guillard1999behavior,
  title={On the behavior of upwind schemes in the low Mach number limit},
  author={Guillard, Herv{\'e} and Viozat, Claude},
  journal={Computers \& Fluids},
  volume={28},
  number={1},
  pages={63--86},
  year={1999},
  publisher={Elsevier}
}

@article{verwer1986conservative,
  title={Conservative and Non. conservative Schemes for the Solution of the Nonlinear Schrodinger Equation},
  author={Verwer, JG},
  journal={IMA journal of Numerical analysis},
  volume={6},
  pages={25--42},
  year={1986}
}

@article{castro2006well,
  title={Well-balanced high-order finite volume methods for systems of balance laws},
  author={Castro, Mariano and Par{\'e}s, Carlos},
  journal={Journal of Scientific Computing},
  volume={27},
  number={1-3},
  pages={107--131},
  year={2006},
  publisher={Springer}
}

@article{dal1995definition,
  title={Definition and weak stability of nonconservative products},
  author={Dal Maso, Gianni and LeFloch, Philippe G and Murat, Fran{\c{c}}ois},
  journal={Journal de Math{\'e}matiques Pures et Appliqu{\'e}es},
  volume={74},
  number={6},
  pages={483--548},
  year={1995},
  publisher={Elsevier}
}

@article{li2024path,
  title={A Path-Conservative ADER Discontinuous Galerkin Method for Non-Conservative Hyperbolic Systems: Applications to Shallow Water Equations},
  author={Li, Yujie and Zhang, Jiahui and Xia, Yinhua and Xu, Yan},
  journal={Mathematics},
  volume={12},
  number={16},
  pages={2601},
  year={2024},
  publisher={MDPI},
  doi={10.3390/math12162601},
  url={https://www.mdpi.com/2227-7390/12/16/2601}
}

@article{fjordholm2012accurate,
  title={Accurate numerical discretizations of non-conservative hyperbolic systems},
  author={Fjordholm, Ulrik Skre and Mishra, Siddhartha},
  journal={ESAIM: Mathematical Modelling and Numerical Analysis},
  volume={46},
  number={1},
  pages={187--206},
  year={2012},
  publisher={EDP Sciences},
  doi={10.1051/m2an/2011044},
  url={https://www.esaim-m2an.org/articles/m2an/abs/2012/01/m2an110044/m2an110044.html}
}

@book{LeVeque2002,
  author    = {R. J. LeVeque},
  title     = {Finite Volume Methods for Hyperbolic Problems},
  publisher = {Cambridge University Press},
  address   = {Cambridge, UK},
  year      = {2002},
  isbn      = {9780521815876}
}

@article{Baer1986,
  author    = {M. R. Baer and J. W. Nunziato},
  title     = {A two-phase mixture theory for the deflagration-to-detonation transition (DDT) in reactive granular materials},
  journal   = {International Journal of Multiphase Flow},
  year      = {1986},
  volume    = {12},
  number    = {6},
  pages     = {861--889},
  doi       = {10.1016/0301-9322(86)90033-9}
}

@book{Godunov2003,
  author    = {S. K. Godunov and E. I. Romenskii},
  title     = {Elements of Continuum Mechanics and Conservation Laws},
  publisher = {Springer},
  address   = {Springer},
  year      = {2003},
  isbn      = {9781402011981}
}

@article{Kemm2002,
  author    = {F. Kemm and D. Kröner and C. D. Munz and T. Schnitzer and M. Wesenberg},
  title     = {Hyperbolic divergence cleaning for the MHD equations},
  journal   = {Journal of Computational Physics},
  year      = {2002},
  volume    = {175},
  number    = {2},
  pages     = {645--673},
  doi       = {10.1006/jcph.2001.6961}
}

@article{aw2000resurrection,
  author    = {Aw, A. and Rascle, M.},
  title     = {Resurrection of ``second order'' models of traffic flow},
  journal   = {SIAM Journal on Applied Mathematics},
  volume    = {60},
  number    = {3},
  pages     = {916--938},
  year      = {2000},
  doi       = {10.1137/S0036139997332099}
}

@book{celia1990general,
  title={A general mass-conservative numerical solution for the unsaturated flow equation},
  author={Celia, Michael A. and Bouloutas, Elias T. and Zarba, Ronald L.},
  journal={Water Resources Research},
  volume={26},
  number={7},
  pages={1483--1496},
  address = {AGU},
  year={1990},
  publisher={Wiley},
  doi={10.1029/WR026i007p01483}
}

@book{tanner2013rheology,
  title={Rheology: An Historical Perspective},
  author={Tanner, Roger I. and Walters, Ken},
  year={2013},
  publisher={Elsevier}
}

@book{goldstein2002classical,
  title={Classical Mechanics},
  author={Goldstein, Herbert and Poole, Charles and Safko, John},
  edition={3rd},
  address = {University of Toronto}
  year={2002},
  publisher={Addison-Wesley},
  isbn={9780201657029}
}

@article{berthon2007augmented,
  title={Augmented Roe schemes for nonconservative hyperbolic systems},
  author={Berthon, Christophe and Coquel, Frédéric},
  journal={SIAM Journal on Scientific Computing},
  volume={29},
  number={6},
  pages={2370--2397},
  year={2007}
}

@article{greenberg1996well,
  title={Well-balanced schemes for conservation laws with source terms},
  author={Greenberg, J.M. and LeRoux, A.Y.},
  journal={SIAM Journal on Numerical Analysis},
  volume={33},
  number={1},
  pages={1--16},
  year={1996}
}

@article{harten1983upstream,
  title={Upstream differencing and Godunov-type schemes for hyperbolic conservation laws},
  author={Harten, Ami},
  journal={SIAM Review},
  volume={25},
  number={1},
  pages={35--61},
  year={1983}
}

@article{berger1989local,
  title={Local adaptive mesh refinement for shock hydrodynamics},
  author={Berger, Marsha J. and Colella, Phillip},
  journal={Journal of Computational Physics},
  volume={82},
  number={1},
  pages={64--84},
  year={1989}
}

@inproceedings{ray2021data,
  title={A data-driven approach to predict artificial viscosity in high-order solvers},
  author={Ray, Deep},
  booktitle={2022 Spring Central Sectional Meeting},
  year={2021},
  organization={AMS}
}

@article{bois2023optimal,
  title={An Optimal Control Deep Learning Method to Design Artificial Viscosities for Discontinuous Galerkin Schemes},
  author={Bois, Léo and Franck, Emmanuel and Navoret, Laurent and Vigon, Vincent},
  journal={arXiv preprint arXiv:2309.11795},
  year={2023}
}

@article{coutinho2023physics,
  title={Physics-informed neural networks with adaptive localized artificial viscosity},
  author={Coutinho, Emilio Jose Rocha and Dall'Aqua, Marcelo and McClenny, Levi and Zhong, Ming and Braga-Neto, Ulisses and Gildin, Eduardo},
  journal={Journal of Computational Physics},
  volume={489},
  pages={112265},
  year={2023},
  publisher={Elsevier}
}

@article{neelan2024physics,
  title={Physics-informed neural networks and higher-order high-resolution methods for resolving discontinuities and shocks: A comprehensive study},
  author={Neelan, Arun Govind and Krishna, G Sai and Paramanantham, Vinoth},
  journal={Journal of Computational Science},
  volume={83},
  pages={102466},
  year={2024},
  publisher={Elsevier}
}

@article{govind2022higher,
  title={Higher-order slope limiters for Euler equation},
  author={Govind Neelan, Arun and Nair, Manoj T},
  journal={Journal of Applied and Computational Mechanics},
  volume={8},
  number={3},
  pages={904--917},
  year={2022},
  publisher={Shahid Chamran University of Ahvaz}
}

@article{neelan2023efficient,
  title={An efficient three-level weighted essentially non-oscillatory scheme for hyperbolic equations},
  author={Neelan, A Arun Govind and Chandran, R Jishnu and Diaz, Manuel A and B{\"u}rger, Raimund},
  journal={Computational and Applied Mathematics},
  volume={42},
  number={2},
  pages={70},
  year={2023},
  publisher={Springer}
}

@article{neelan2021three,
  title={Three-level order-adaptive weighted essentially non-oscillatory schemes},
  author={Neelan, A Arun Govind and Nair, Manoj T and B{\"u}rger, Raimund},
  journal={Results in Applied Mathematics},
  volume={12},
  pages={100217},
  year={2021},
  publisher={Elsevier}
}

@article{drozda2023learning,
  title={Learning an optimised stable Taylor-Galerkin convection scheme based on a local spectral model for the numerical error dynamics},
  author={Drozda, Luciano and Mohanamuraly, Pavanakumar and Cheng, Lionel and Lapeyre, Corentin and Daviller, Guillaume and Realpe, Yuval and Adler, Amir and Staffelbach, Gabriel and Poinsot, Thierry},
  journal={Journal of Computational Physics},
  volume={493},
  pages={112430},
  year={2023},
  publisher={Elsevier}
}

@inproceedings{dubey2021learning,
  title={Learning numerical viscosity using artificial neural regression network},
  author={Dubey, Ritesh Kumar and Gupta, Anupam and Jayswal, Vikas Kumar and Pandey, Prashant Kumar},
  booktitle={Computational Sciences-Modelling, Computing and Soft Computing: First International Conference, CSMCS 2020, Kozhikode, Kerala, India, September 10-12, 2020, Revised Selected Papers 1},
  pages={42--55},
  year={2021},
  organization={Springer}
}

@article{dwivedi2019distributed,
  title={Distributed physics informed neural network for data-efficient solution to partial differential equations},
  author={Dwivedi, Vikas and Parashar, Nishant and Srinivasan, Balaji},
  journal={arXiv preprint arXiv:1907.08967},
  year={2019}
}

@article{SHUKLA2021110683,
  title={Parallel physics-informed neural networks via domain decomposition},
  author={Shukla, Khemraj and Jagtap, Ameya D and Karniadakis, George Em},
  journal={Journal of Computational Physics},
  volume={447},
  pages={110683},
  year={2021},
  publisher={Elsevier}
}

@article{JAGTAP2020113028,
  title={Conservative physics-informed neural networks on discrete domains for conservation laws: Applications to forward and inverse problems},
  author={Jagtap, Ameya D and Kharazmi, Ehsan and Karniadakis, George Em},
  journal={Computer Methods in Applied Mechanics and Engineering},
  volume={365},
  pages={113028},
  year={2020},
  publisher={Elsevier}
}

@article{CiCP-28-5,
  title={Extended physics-informed neural networks (XPINNs): A generalized space-time domain decomposition based deep learning framework for nonlinear partial differential equations},
  author={Jagtap, Ameya D and Karniadakis, George Em},
  journal={Communications in Computational Physics},
  volume={28},
  number={5},
  year={2020},
  publisher={Brown Univ., Providence, RI (United States)}
}

@article{kharazmi2019variationalphysicsinformedneuralnetworks,
  title={hp-VPINNs: Variational physics-informed neural networks with domain decomposition},
  author={Kharazmi, Ehsan and Zhang, Zhongqiang and Karniadakis, George Em},
  journal={Computer Methods in Applied Mechanics and Engineering},
  volume={374},
  pages={113547},
  year={2021},
  publisher={Elsevier}
}

@article{Berrone_2024,
  title={Meshfree Variational-Physics-Informed Neural Networks (MF-VPINN): An Adaptive Training Strategy},
  author={Berrone, Stefano and Pintore, Moreno},
  journal={Algorithms},
  volume={17},
  number={9},
  pages={415},
  year={2024},
  publisher={MDPI}
}

@inproceedings{rajvanshi2024integral,
  title={Integral pinns for hyperbolic conservation laws},
  author={Rajvanshi, Manvendra P and Ketcheson, David I},
  booktitle={ICLR 2024 Workshop on AI4DifferentialEquations In Science},
  year={2024}
}

@article{moseley2023finite,
  title={Finite basis physics-informed neural networks (FBPINNs): a scalable domain decomposition approach for solving differential equations},
  author={Moseley, Ben and Markham, Andrew and Nissen-Meyer, Tarje},
  journal={Advances in Computational Mathematics},
  volume={49},
  number={4},
  pages={62},
  year={2023},
  publisher={Springer}
}

@article{xiang2022self,
  title={Self-adaptive loss balanced physics-informed neural networks},
  author={Xiang, Zixue and Peng, Wei and Liu, Xu and Yao, Wen},
  journal={Neurocomputing},
  volume={496},
  pages={11--34},
  year={2022},
  publisher={Elsevier}
}

@article{Lu_2021,
  title={Learning nonlinear operators via DeepONet based on the universal approximation theorem of operators},
  author={Lu, Lu and Jin, Pengzhan and Pang, Guofei and Zhang, Zhongqiang and Karniadakis, George Em},
  journal={Nature machine intelligence},
  volume={3},
  number={3},
  pages={218--229},
  year={2021},
  publisher={Nature Publishing Group UK London}
}

@article{patel2024variationally,
  title={Variationally mimetic operator networks},
  author={Patel, Dhruv and Ray, Deep and Abdelmalik, Michael RA and Hughes, Thomas JR and Oberai, Assad A},
  journal={Computer Methods in Applied Mechanics and Engineering},
  volume={419},
  pages={116536},
  year={2024},
  publisher={Elsevier}
}

@article{li2024physics,
  title={Physics-informed neural operator for learning partial differential equations},
  author={Li, Zongyi and Zheng, Hongkai and Kovachki, Nikola and Jin, David and Chen, Haoxuan and Liu, Burigede and Azizzadenesheli, Kamyar and Anandkumar, Anima},
  journal={ACM/JMS Journal of Data Science},
  volume={1},
  number={3},
  pages={1--27},
  year={2024},
  publisher={ACM New York, NY}
}

@article{li2020fourier,
  title={Fourier neural operator for parametric partial differential equations},
  author={Li, Zongyi and Kovachki, Nikola and Azizzadenesheli, Kamyar and Liu, Burigede and Bhattacharya, Kaushik and Stuart, Andrew and Anandkumar, Anima},
  journal={arXiv preprint arXiv:2010.08895},
  year={2020}
}

@article{mao2020physics,
  title={Physics-informed neural networks for high-speed flows},
  author={Mao, Zhiping and Jagtap, Ameya D and Karniadakis, George E},
  journal={Journal of Computational Physics},
  volume={412},
  pages={109439},
  year={2020},
  publisher={Elsevier}
}

@article{wang2021learning,
  title={Learning shock waves in multiphase flows using physics-informed neural networks},
  author={Wang, Renzheng and Wang, Yang and Karniadakis, George E},
  journal={Physics of Fluids},
  volume={33},
  number={8},
  pages={087114},
  year={2021},
  publisher={AIP Publishing}
}

@article{wassing2024physics,
  title={Physics-Informed Neural Networks for Transonic Flows around an Airfoil},
  author={Wassing, Simon and Langer, Stefan and Bekemeyer, Philipp},
  journal={arXiv preprint arXiv:2408.17364},
  year={2024}
}

@article{liang2024continuous,
  title={Continuous and discontinuous compressible flows in a converging--diverging channel solved by physics-informed neural networks without exogenous data},
  author={Liang, Hong and Song, Zilong and Zhao, Chong and Bian, Xin},
  journal={Scientific Reports},
  volume={14},
  number={1},
  pages={3822},
  year={2024},
  publisher={Nature Publishing Group UK London}
}

@article{lee2025physics,
  title={Physics informed neural networks for fluid flow analysis with repetitive parameter initialization},
  author={Lee, Jongmok and Shin, Seungmin and Kim, Taewan and Park, Bumsoo and Choi, Ho and Lee, Anna and Choi, Minseok and Lee, Seungchul},
  journal={Scientific Reports},
  volume={15},
  number={1},
  pages={1--16},
  year={2025},
  publisher={Nature Publishing Group}
}

@article{fedkiw2002general,
  title={A general technique for eliminating spurious oscillations in conservative schemes for multiphase and multispecies Euler equations},
  author={Fedkiw, Ronald and Liu, X-D and Osher, Stanley},
  journal={International Journal of Nonlinear Sciences and Numerical Simulation},
  volume={3},
  number={2},
  pages={99--106},
  year={2002},
  publisher={De Gruyter}
}

@article{arun2018hyperbolic,
  title={Hyperbolic Runge--Kutta method using evolutionary algorithm},
  author={Arun Govind Neelan, A and Nair, Manoj T},
  journal={Journal of Computational and Nonlinear Dynamics},
  volume={13},
  number={10},
  pages={101003},
  year={2018},
  publisher={American Society of Mechanical Engineers}
}

@article{neelan2025higher,
  title={Higher-order conservative discretizations on arbitrarily varying non-uniform grids},
  author={Neelan, A Arun Govind and B{\"u}rger, Raimund and Nair, Manoj T and Rathan, Samala},
  journal={Computational and Applied Mathematics},
  volume={44},
  number={1},
  pages={27},
  year={2025},
  publisher={Springer}
}

@article{pyclaw-sisc,
  Author = {Ketcheson, David I. and Mandli, Kyle T. and Ahmadia, Aron J. and Alghamdi, Amal and {Quezada de Luna}, Manuel and Parsani, Matteo and Knepley, Matthew G. and Emmett, Matthew},
  Journal = {SIAM Journal on Scientific Computing},
  Month = nov,
  Number = {4},
  Pages = {C210--C231},
  Title = {{PyClaw: Accessible, Extensible, Scalable Tools for Wave Propagation Problems}},
  Volume = {34},
  Year = {2012}
}

@incollection{leveque-george-2008,
  author = {LeVeque, Randall J. and George, David L.},
  title = {High-Resolution Finite Volume Methods for the Shallow Water Equations with Bathymetry and Dry States},
  booktitle = {Advances in Coastal and Ocean Engineering},
  publisher = {World Scientific},
  year = {2008},
  volume = {10},
  pages = {43--73},
  doi = {10.1142/9789812814043_0002}
}

@article{lax1954weak,
  title={Weak solutions of nonlinear hyperbolic equations and their numerical computation},
  author={Lax, Peter D.},
  journal={Communications on Pure and Applied Mathematics},
  volume={7},
  number={1},
  pages={159--193},
  year={1954},
  publisher={Wiley Online Library}
}

@article{harten1983high,
  title={High resolution schemes for hyperbolic conservation laws},
  author={Harten, Ami},
  journal={Journal of Computational Physics},
  volume={49},
  number={3},
  pages={357--393},
  year={1983},
  publisher={Elsevier}
}

@article{hou1994non,
  title={Why nonconservative schemes converge to wrong shock solutions: Error analysis},
  author={Hou, Thomas Y. and LeFloch, Philippe G.},
  journal={Mathematics of Computation},
  volume={62},
  number={206},
  pages={497--530},
  year={1994}
}

@book{leveque1992numerical,
  title={Numerical Methods for Conservation Laws},
  author={LeVeque, Randall J},
  year={1992},
  publisher={Birkh{\"a}user Basel}
}

@article{lax1960systems,
  title={Systems of conservation laws},
  author={Lax, Peter D and Wendroff, Burton},
  journal={Communications on Pure and Applied Mathematics},
  volume={13},
  number={2},
  pages={217--237},
  year={1960},
  publisher={Wiley Online Library}
}

@article{pares2006numerical,
  title={Numerical methods for nonconservative hyperbolic systems: a theoretical framework},
  author={Par{\'e}s, Carlos},
  journal={SIAM Journal on Numerical Analysis},
  volume={44},
  number={1},
  pages={300--321},
  year={2006},
  publisher={SIAM}
}

@article{von1950method,
  title={A method for the numerical calculation of hydrodynamic shocks},
  author={Von Neumann, John and Richtmyer, Robert D},
  journal={Journal of Applied Physics},
  volume={21},
  number={3},
  pages={232--237},
  year={1950},
  publisher={American Institute of Physics}
}

@inproceedings{persson2006sub,
  title={Sub-cell shock capturing for discontinuous Galerkin methods},
  author={Persson, Per-Olof and Peraire, Jaime},
  booktitle={44th AIAA Aerospace Sciences Meeting and Exhibit},
  pages={112},
  year={2006}
}

@article{tadmor2003entropy,
  title={Entropy stability theory for difference approximations of nonlinear conservation laws and related time-dependent problems},
  author={Tadmor, Eitan},
  journal={Acta Numerica},
  volume={12},
  pages={451--512},
  year={2003},
  publisher={Cambridge University Press}
}

@article{castro2013entropy,
  title={Entropy conservative and entropy stable schemes for nonconservative hyperbolic systems},
  author={Castro, Manuel J and Fjordholm, Ulrik S and Mishra, Siddhartha and Par{\'e}s, Carlos},
  journal={SIAM Journal on Numerical Analysis},
  volume={51},
  number={3},
  pages={1371--1391},
  year={2013},
  publisher={SIAM}
}

@article{KUMAR2026106975,
title = {A robust data-free physics-informed neural network for compressible flows with shocks},
journal = {Computers \& Fluids},
volume = {308},
pages = {106975},
year = {2026},
issn = {0045-7930},
doi = {https://doi.org/10.1016/j.compfluid.2026.106975},
url = {https://www.sciencedirect.com/science/article/pii/S0045793026000174},
author = {Prashant Kumar and Rajesh Ranjan},
}

@article{munoz2007godunov,
  title={Godunov method for nonconservative hyperbolic systems},
  author={Mu{\~n}oz-Ruiz, Mar{\'\i}a Luz and Par{\'e}s, Carlos},
  journal={ESAIM: Mathematical Modelling and Numerical Analysis},
  volume={41},
  number={1},
  pages={169--185},
  year={2007},
  publisher={EDP Sciences}
}

@article{abgrall2010comment,
  title={A comment on the computation of non-conservative products},
  author={Abgrall, R{\'e}mi and Karni, Smadar},
  journal={Journal of Computational Physics},
  volume={229},
  number={8},
  pages={2759--2763},
  year={2010},
  publisher={Elsevier}
}
\end{document}